\documentclass{jfm}
\usepackage{graphicx}
\usepackage{epstopdf, epsfig}
\usepackage{hyperref}
\usepackage{amsmath}
\usepackage{amssymb}
\usepackage{placeins}
\usepackage{color,soul}
\usepackage{enumerate}
\usepackage{bm}
\usepackage{natbib}

\newcommand{\reva}[1]{{\color{black} #1}}
\newcommand{\revb}[1]{{\color{black} #1}}
\newcommand{\revc}[1]{{\color{black} #1}}
\newcommand{\revd}[1]{{\color{black} #1}}
\newcommand{\reve}[1]{{\color{black} #1}}

\title{Elasto-inertial transitions in viscoelastic flows through cylinder arrays}
\author{J. R. C. King\aff{1}
  \corresp{\email{jack.king@manchester.ac.uk}},
  H. M. Broadley\aff{2}, \and 
  M. Beneitez\aff{1}}

\affiliation{\aff{1}Department of Mechanical and Aerospace Engineering, The University of Manchester, Manchester, UK\\
\aff{2}School of Engineering, University of Liverpool, Liverpool, UK
}

\begin{document}

\maketitle
\begin{abstract}

%% Big picture
For dilute solutions of polymers, chaotic flow states can occur at lower Reynolds numbers than required for inertial turbulence in Newtonian fluids, offering the potential for increased mixing efficiency. These states may be promoted by the flow geometry, and in recent years, porous media have gained attention as a promising setting in which viscoelastic instabilities may be exploited, although studies have primarily been in the creeping flow regime. Cylinder arrays serve as a prototypical porous media, giving a controlled setting in which to investigate flow dynamics. Here we explore the transition to elasto-inertial turbulence (EIT) in cylinder arrays via detailed \reva{two-dimensional} numerical simulations \reva{for $Re\in\left[10,500\right]$ and $Wi\in\left[0,320\right]$}. With increasing elasticity, EIT is reached via an initial sub-critical saddle-node bifurcation from the Newtonian state and then follows a series of supercritical bifurcations, in a Ruelle-Takens-Newhouse route to chaos. This transition is driven by the interaction between vortex shedding in cylinder wakes, and the bulk flow between cylinders. Within the EIT regime, we observe an interaction between slow dynamics in cylinder wakes, and fast dynamics in channels between cylinders, leading to two distinct slopes in the energy spectra. At low Reynolds numbers arrowhead structures are present, but these are suppressed at higher inertia. In the present configuration, we find no direct connection between EIT and purely elastic instabilities.

%% Sentence putting implications in perspective?

\end{abstract}

\maketitle
%end front matter

\section{Introduction}\label{sec:intro}

%% Motivation, big picture etc.
Flows in porous media are ubiquitous, occurring \revb{across a range of scales in both} industry and nature, from packed bed catalysers, heat exchangers, nuclear pebble-bed reactors, and enhanced oil-recovery processes, to near surface atmospheric flows and sub-surface magma convection~\citep{wood_2020}. Many porous media flows involve polymer solutions with viscoelastic rheology. Even in simple geometries, such fluids can exhibit markedly different dynamics from their Newtonian counterparts.

%% Intro to ET and EIT.
Elastic turbulence (ET) and elasto-inertial turbulence (EIT) are two (related) phenomena present only in viscoelastic fluids. Elastic turbulence (ET) is a chaotic flow state first formally described a quarter of a century ago~\citep{groisman_2000}, which can exist even in the limit of vanishing inertia. Elasto-inertial turbulence (EIT)~\citep{samanta_2013} is a chaotic flow state with similar characteristics (i.e. energy spectra) but which occurs when both elastic and inertial effects are important. In these regimes polymer solutions can exhibit chaotic dynamics with fluctuations over a broad range of length- and time-scales, at flow rates where Newtonian fluids would remain laminar. ET and EIT have the potential to increase mixing and emulsification~\citep{poole_2012} and heat transfer~\citep{traore_2015,abed_2016}, and a comprehensive understanding of their mechanisms and transition has been the subject of much research amongst the community over the past two decades. Much of this research has focused on the canonical geometries of Taylor-Couette and channel flows, where differing instability mechanisms have been identified as the precursor to chaotic dynamics. In the curved geometries of Taylor-Couette flows, a purely elastic instability exists due to hoop-stresses~\citep{shaqfeh_1996}. For channel flows, a direct sub-critical transition has been proposed~\citep{morozov_2007}, and consequently much work has focused on exact coherent structures, with the \emph{arrowhead}~\citep{page_2020,dubief_2022} structure increasingly thought to be key. \revd{Note the same structure is referred to as a \emph{narwhal} by~\citep{morozov_2022} and collaborators.} For reviews of ET and EIT, we refer the interested reader to~\cite{steinberg_2021,dubief_2023,datta_2022a}.

% Browne, datta work
Recently, the potential of viscoelastic flows in porous media to promote mixing, homogenization and reaction has been highlighted in a series of experimental works~\citep{browne_2021,browne_2023,browne_2024}, showing high-quality imaging of polymer flows through disordered porous media. Whilst this setting has practical relevance, the geometric complexity is such that a fundamental understanding of the pore-scale dynamics remains elusive\revb{. Aside from the recent work by~\cite{zhu_2024}, there has been very limited investigation of inertial flows in porous geometries}, and a number of open questions remain: How are ET and EIT states connected in confined geometries? What are the mechanisms controlling transition? Are instabilities hoop-stress related, or arrowhead-initiated? Does EIT stem from a purely elastic instability in confined geometries?

%% cylinder arrays
To investigate these questions we simplify the problem from disordered porous media to cylinder arrays. This provides a useful canonical geometry in which to study flow dynamics and transition pathways~\revb{in a controlled environment. I}n recent years \revb{cylinder arrays have been used} as a test-bed to explore symmetry breaking and transition in inertial turbulence~\citep{srikanth_2025}, the existence of large super-pore scale structures~\citep{jin_2015}\revb{, and transitions in viscoelastic flows~\citep{kumar_2021,kumar_2023,walkama_2020,haward_2021}}.

%% Previous work on this specific topic
In the field of viscoelastic flows, early experimental work on polymer flows past cylinder arrays predated the formal discovery of ET, though increasing unsteadiness and asymmetry with increasing flow resistance at higher elasticities was observed~\citep{chmie_1993,khomami_1997}, which we now see to be the hallmarks of ET. Around the same time, numerous numerical investigations were conducted of similar problems, although in some cases subject to the assumptions of symmetry~\citep{liu_1998} or steadiness~\citep{alcocer_2002}. ~\cite{talwar_1995} simulated the flow of polymer solutions around arrays of cylinders, both with inertia and in a creeping flow limit. The flow was assumed steady, and in some cases symmetry was imposed, but they were able to infer the existence of a temporal instability - as reported earlier by~\cite{chmie_1993}. Meanwhile, simulations by~\cite{mckinley_1993} of a single cylinder in a channel showed the onset of a wake instability, with the mechanism linked to streamline curvature due to confinement - a mechanism also likely at play in cylinder arrays. ~\cite{oliveira_2001} performed simulations of the time-dependent vortex shedding behind a cylinder, finding the shedding frequency was reduced by elasticity, and region of wake elongated. 

More recently, experiments at the microscale of viscoelastic flow past cylinders~\citep{galindo_2014,ribeiro_2014} and linear cylinder arrays~\citep{kenney_2013,shi_2015,shi_2016} found the development of chaotic flow originating from instabilities in the trailing edge stagnation points, whilst experimental evidence of symmetry breaking for a single cylinder was found by~\cite{hopkins_2020}.
~\cite{walkama_2020} experimentally studied viscoelastic flows through cylinder arrays, finding that disorder suppressed the transition to chaotic dynamics. In experiments on a similar configuration - but with the flow direction rotated relative to the cylinder array by 30 degrees -~\cite{haward_2021} found a contradictory result - that disorder promoted chaos. This was explained by the different orientation, with the key inference being that the exposure of stagnation points to the flow drives chaotic dynamics, a finding in keeping with the stagnation point instabilities in linear arrays found by~\cite{shi_2016}. 

%% What we do.
In this work we numerically study \reva{two-dimensional }flows of dilute polymer solutions through cylinder arrays at non-negligible inertia \reva{($Re\ge10$)} \reva{We use a high-order mesh-free framework~\cite{king_2024a}, and following the approach widely taken in the mesh-free community (e.g.~\cite{vq_ellero_2012}), use a weakly compressible formulation, whilst ensuring the (small) influence of compressibility plays a negligible role on the dynamics (see Appendix~\ref{sec:conv}).} Our aim is to identify the pathways to transition in such settings, and to understand the origin of EIT in cylinder arrays.

%% Numerics are tricky
Numerical studies of such geometries are challenging. ET/EIT flows are characterised by thin, dynamically evolving sheets of highly stretched polymers, requiring highly accurate (low-dissipation) methods to resolve. Most numerical methods suitable for such complex geometries (such as unstructured finite volume methods) are restricted to low-order and are overly dissipative for the dynamics of ET/EIT. Pseudo-spectral methods with immersed boundaries have been used to study similar geometries~\citep{zhu_2024}, with a focus on arrowhead structures, although in such methods accurate imposition of boundary conditions remains a challenge. \revc{In that study, transition to EIT in a Cartesian lattice was investigated by artificially triggering arrowhead structures, with transition to a chaotic state observed at much smaller $Re$ than the Newtonian case.} A further challenge arises in these flows due to the recently discovered polymer-diffusive instability (PDI)~\citep{beneitez_2023}; a linear instability arising in wall-bounded flows due to the presence of polymer diffusion, whether explicitly added to provide numerical stability, or implicitly included in the numerical method \cite{beneitez_2025}. The most unstable wavelength of PDI scales with the square root of diffusivity, and hence as diffusivity reduces, PDI can grow at scales below the continuum limit - outside the validity of the constitutive models. With hindsight, the signs of PDI \reva{- near-wall structures with small-scale features in the streamwise direction -} may be seen in numerous works. For example: the Smoothed Particle Hydrodynamics simulations of linear cylinder arrays by~\cite{grilli_2013} \reva{(Figure 4, and the SI);} the hybrid Lattice-Boltzmann/finite-difference simulations of individual pores by~\cite{dzanic_2023} \reva{(Figures 2, 5, 9, 11 and 12)}; the finite-volume simulations of~\cite{kumar_2021} \reva{(Figures 5 and 9),} \reva{~\cite{kumar_2021b} (Figures 2, 3 and 4), and~\cite{kumar_2023} (Figure 4b)}; \reva{finite difference simulations of channel flow by~\cite{giulio_2024} (Figure 1);} and previous research by the lead author of the present work~\citep{king_2024a} \reva{(Figures 13 and 14)}. PDI can be sufficient to trigger a transition to an ET-like state, even where the base flow (in the absence of polymer diffusivity) is linearly stable~\citep{beneitez2024transition}. Whilst chaotic dynamics in the configurations simulated by the above authors may exist, the observed dynamics are likely to have been influence, or at least triggered, by PDI. \revc{Given the most unstable wavelength of PDI tends to zero with vanishing diffusivity, we believe it to be an unphysical consequence of the continuum constitutive models used.} Developing robust stabilisation strategies for numerical simulations of viscoelastic flows which do not exhibit PDI is an important challenge for computational rheologists. In lieu of meeting this challenge, in the present work \revb{we have verified that PDI is not present in our simulations and does not underpin the observed dynamics.}

% Layout
The remainder of this paper is structured as follows. In~\S~\ref{pc} we set out details of the geometry and governing equations of the system we study.~\S~\ref{nm} provides details of the numerical methods we use. In~\S~\ref{results} we present simulation results, alongside analysis and discussion of the transition mechanisms at play. In~\S~\ref{sec:conc} we provide concluding remarks, and thoughts on future research in this area.

\section{Problem configuration}\label{pc}

We consider isothermal flows of polymer solutions, in which the polymers are assumed to obey a simplified Phan-Thien-Tanner (sPTT) model~\citep{ptt_1977}. The flow is governed by conservation equations for mass and momentum, alongside an evolution equation for the conformation tensor. These are given in dimensionless form as
\begin{equation}\frac{\partial\rho}{\partial{t}}+\frac{\partial\rho{u}_{j}}{\partial{x}_{j}}=0,\label{eq:mass}\end{equation}
\begin{equation}\frac{\partial\rho{u}_{i}}{\partial{t}}+\frac{\partial\rho{u}_{i}{u}_{j}}{\partial{x}_{j}}=-\frac{\partial{p}}{\partial{x_{i}}}+\frac{\beta}{Re}\frac{\partial^{2}u_{i}}{\partial{x}_{j}\partial{x}_{j}}+\frac{\left(1-\beta\right)}{ReWi}\frac{{c}_{ji}}{\partial{x}_{j}}+F_{0}\delta_{i1}\label{eq:mom},\end{equation}
\begin{equation}\frac{\partial{c}_{ij}}{\partial{t}}+u_{k}\frac{\partial{c}_{ij}}{\partial{x}_{k}}-\frac{\partial{u}_{i}}{\partial{x}_{k}}c_{kj}-\frac{\partial{u}_{j}}{\partial{x}_{k}}c_{ik}=-\frac{1-3\varepsilon+\varepsilon{c}_{kk}}{Wi}\left(c_{ij}-\delta_{ij}\right)+\kappa\frac{\partial^{2}c_{ij}}{\partial{x}_{k}\partial{x}_{k}},\label{eq:cte}\end{equation}
in which $\rho$ is density, $u_{i}$ is the velocity, $c_{ij}$ is the conformation tensor, and $p$ is the pressure. $F_{0}\delta_{i1}$ is a dimensionless pressure gradient imposed to drive the flow. The system of governing equations is closed with a barotropic equation of state
\begin{equation}p=\frac{\rho}{Ma^{2}}.\end{equation}

The dimensionless quantities governing the system are the Reynolds number $Re$, the Mach number $Ma$, the Weissenberg number $Wi$, the sPTT nonlinearity parameter $\varepsilon$, the viscosity ratio $\beta$, and the dimensionless diffusivity $\kappa$. 
The system~\eqref{eq:mass} to~\eqref{eq:cte} is subject to no slip and zero normal diffusi\reva{ve flux} conditions on solid boundaries:
\begin{equation}u_{i}=0;\qquad{n}_{k}\frac{\partial{c}_{ij}}{\partial{x}_{k}}=0\end{equation}
where $n_{k}$ is the unit vector normal to the boundary.

\begin{figure}
\includegraphics[width=0.43\textwidth]{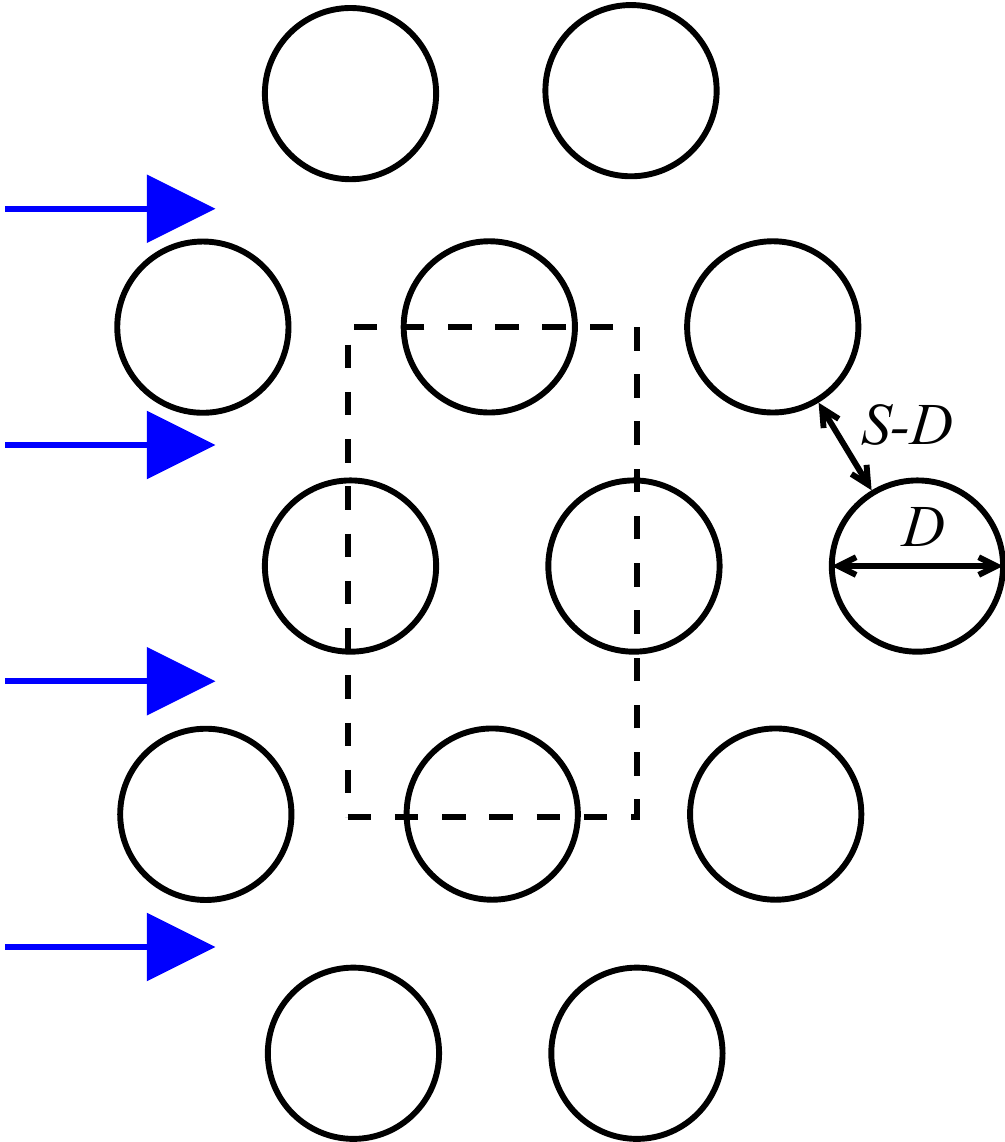}
\includegraphics[width=0.49\textwidth]{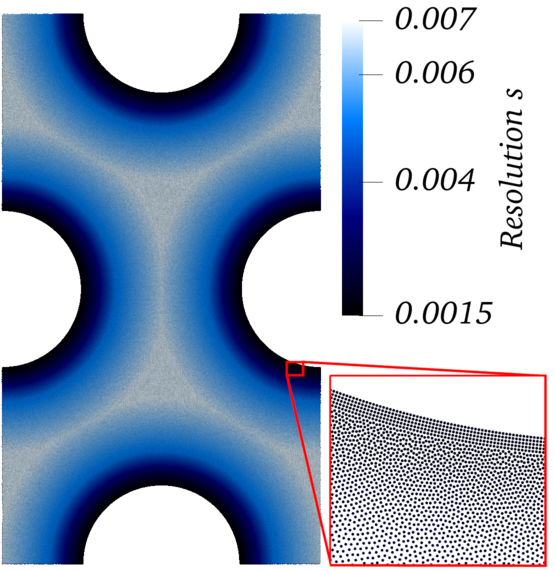}
\caption{Left panel: A schematic of the flow configuration. The computational domain is the minimal periodic unit of the cylinder array shown by a dashed line. Right panel: isocontours of resolution, with the inset illustrating a typical node distribution near the boundary.\label{fig:hex_array}}
\end{figure}

We consider flow through a doubly-periodic unit cell of a hexagonal lattice of cylinders, as shown in the left panel of figure~\ref{fig:hex_array}. The flow is made dimensionless by the cylinder diameter $D$ and the velocity $U=\dot{Q}/A$, where $A$ is the cross sectional area of the domain normal to the flow direction, and $\dot{Q}$ is the volumetric flow rate in the Newtonian limit for a given viscosity and value of $F_{0}$. The porosity $\phi$ is linked to the cylinder spacing $S$ by $\phi=1-\pi/\left(2\sqrt{3}S^{2}\right)$, and throughout this work we set $S=2$, giving a porosity of $\phi=1-\pi/8\sqrt{3}\approx0.773$. Except where otherwise specified, we assume the flow is planar, and solve the two-dimensional problem. Where we solve the three-dimensional problem (only for Newtonian baselines), periodicity of \revc{up to length $4D$} is imposed in the third dimension. 

In this work we fix $\beta=0.9$, representing dilute polymer solutions, $\varepsilon=10^{-3}$ and $\kappa=10^{-5}$. Except where explicitly stated, we set $Ma=0.01$. Precursor Newtonian simulations at $Re\in\left[10,25,50,100,250,500\right]$ in which the pressure gradient is dynamically adjusted via a PID controller targeting $\dot{Q}=A$, yielding corresponding values of $F_{0}=\left[2.053,0.975,0.561,0.343,0.184,0.109\right]$. For simulations at intermediate values of $Re$, values of $F_{0}$ are obtained via interpolation, assuming a power-law relationship $F_{0}\propto{Re}^{-\alpha}$, with a piecewise-constant exponent. \revb{Note that for any given viscosity, our dimensionless parameters are then defined based on the timescale of the Newtonain flow at that viscosity, in the same way manner as~\cite{zhu_2024}.}

Similar geometries have previously been numerically studied using a weakly compressible SPH formulation~\citep{vq_ellero_2012,grilli_2013}, and Lattice-Boltzmann methods~\citep{gillissen_2013,dzanic_2023} which are inherently weakly compressible. In the former - as in the present case - a barotropic equation of state was used. There, the sound speed was chosen such that $Ma=0.006$, but with the flows under consideration in the creeping regime, a large pressure gradient ($\mathcal{O}\left(10^{2}\right)$) was needed to drive the flow, and the resulting density variations remained under $3\%$. In the present work we only consider $Re\ge{10}$, and consequently the pressure gradient driving the flow is smaller: taking a maximum of $F_{0}=2.053$ at $Re=10$. The maximum density variation in the domain is of the order $SF_{0}Ma^{2}$, and for $Ma=0.01$, density variations remain below $0.1\%$ in all cases (and for larger $Re$, an order of magnitude smaller still). Confirmation that our value of $Ma$ is sufficiently small to have negligible impact on the flow dynamics is provided in Appendix~\ref{sec:conv}.

\reva{Whilst both sPTT and FENE-P models are used for flows of dilute polymer solutions across the community (e.g.~\cite{morozov_2022} and ~\cite{dubief_2022} respectively), the phenomenology of ET/EIT and arrowhead structures is robust across both models. Our choice to use the sPTT model is motivated in part by our experience that it is less prone to PDI than FENE-P, and simulations tend to be less prone to numerical instabilities.} The choice of $\varepsilon$ is consistent with the value used in simulations of ET in channel flows~\citep{morozov_2022} - and provides an equivalent degree of non-linearity to the simulations of FENE-P fluids in~\cite{zhu_2024}. Polymeric diffusivity is essential to regularise~\eqref{eq:cte}, and is widely included explicitly (e.g.~\cite{gillissen_2013,plan_2017,page_2020,morozov_2022,morozov_2025}) or implicitly (e.g.~\cite{dzanic_2023,garg_2025,giulio_2026}). \reva{P}olymeric diffusivity has a physical basis\reva{. When deriving constitutive equations based on the kinetic theory of polymers, it is only due to the assumption of flow homogeneity that the diffusive term does not appear~\citep{elkareh_1989,beris_1994}. In reality, polymer diffusivity is small. S}everal researchers (e.g.~\cite{morozov_2022,nichols_2025}) \reva{justify values of diffusivity based on kinetic theory. We acknowledge that the primary purpose of polymeric diffusivity in the present case is to stabilise our numerical simulations}. The value of $\kappa=10^{-5}$ is smaller than that used in existing studies of ET~\citep{morozov_2022} and comparable with values used for studies of EIT~\citep{page_2020,buza_2022}.

\section{Numerical methods\label{nm}}

%% Numerical challenges
As discussed in~\S~\ref{sec:intro}, simulations of polymer solutions in complex geometries present significant numerical challenges. ET and EIT involve extremely fine filamental stuctures in the polymeric deformation field. Accurate resolution of these structures and their evolving dynamics requires high-fidelity numerical methods, typically with very low numerical diffusion~\citep{yerasi_2024}. Consequently, many studies of ET and EIT use pseudo-spectral methods (e.g.~\cite{morozov_2022,morozov_2025,beneitez_2023}) or high-order finite difference methods (e.g.~\cite{yerasi_2024}), which although extremely accurate, are limited to simple geometries. Finite volume methods, particularly unstructured methods, may be used to simulate flows in complex geometries, but are typically constructed at low order, and the numerical dissipation in such schemes can strongly alter the flow dynamics, and for wall-bounded flows, render the discretized equations linearly unstable~\cite{beneitez_2025}.

In the present work we use a numerical framework which is unique in the viscoelastic community: it is mesh-free, permitting simulations of complex geometries, but high-order with low numerical dissipation, rendering it suitable for simulating the dynamics of ET/EIT. The spatial discretisation is based on the Local Anisotropic Basis Function Method (LABFM)~\cite{king_2020,king_2022}, previously applied to direct numerical simulations of combustion~\cite{king_2024}, and viscoelastic flows~\cite{king_2024a}. We refer readers to these works for complete details of the numerical methods. Here, we provide an overview.

We take a Cholesky decomposition of $c_{ij}=l_{ik}l_{jk}$, and evolve equations for $l_{ij}$ $\forall{i}\ne{j}$ and $\ln{l_{ij}}$ $\forall{i}=j$ to guarantee $c_{ij}$ remains symmetric positive definite~\cite{vaithianathan_2003}. The governing equations are solved on an unstructured set $N_{n}$ of nodes (aka collocation points). The geometry is discretised with uniform nodes along solid boundaries, and a locally uniform distribution near boundaries, following~\cite{king_2022}. The right panel of figure~\ref{fig:hex_array} shows isocontours of resolution, with the inset showing a typical node distribution near the boundary. Internally, the node distribution is generated via a propagating front algorithm~\cite{fornberg_2015a}, in which the resolution $s$ (the local average node spacing) is allowed to vary spatially. In the present work we set $s_{min}=D/600$ at walls, smoothly increasing to $s_{max}=5s_{min}$ far from the cylinder surfaces, and we denote the resolution of a given discretisation by the value of $s_{min}$. We have confirmed resolution independence of our results in Appendix~\ref{sec:conv}. Spatial derivatives are calculated using LABFM, and for details of this procedure, and a comprehensive analysis of the method, we refer the reader to~\cite{king_2020,king_2022}. Derivative approximations are sixth-order accurate internally, reducing to fourth order at non-periodic boundaries. Where we conduct three-dimensional simulations, a uniform node distribution is used in the third (homogeneous) dimension, spatial derivatives in the third dimension are calculated using eighth-order central finite differences. Temporal integration is via an explicit third-order Runge-Kutta scheme with embedded error estimation. The time-step is controlled via a PID controller which ensures errors due to time integration remain below approximately $5\times10^{-6}$. As with other high-order collocated methods, dealiasing is necessary. At each time step we de-alias the solution using an eighth-order filter as in~\cite{king_2022,king_2024a}.

\revc{We note that while our simulations are of a much smaller domain than previous studies (see e.g. \cite{kumar_2023}), the present resolution is an order of magnitude more refined. Given that polymer diffusivity (either explicit or implicitly introduced through numerics) is essential for stability, and the magnitude of the necessary diffusivity scales with the square of the nodal spacing, the minimum diffusivity for previously studied lower resolutions is $\kappa=10^{-3}$. Simulations run at this lower resolution with this larger value of polymeric diffusivity yielded steady flows at parameter values where our more resolved simulations (which are converged -- see Appendix \ref{sec:conv}) were in an EIT state. Whilst large-scale simulations play an important role in identifying super-pore structures in agreement with experiments, the low resolution is insufficient to capture the small-scale dynamics at moderate $Re$ which are the focus of this manuscript.}

\section{Numerical Results}\label{results}

In the following we denote spatially and temporally \revc{volume} averaged quantities by $\left\langle\cdot\right\rangle_{\mathcal{V}}$ and $\left\langle\cdot\right\rangle_{{t}}$ respectively. We denote the volume averaged kinetic energy as $E_{K}=\frac{1}{2}\left\langle\rho{u}_{i}u_{i}\right\rangle_{\mathcal{V}}$, and the volumed averaged conformation tensor trace $E_{E}=\left\langle{c}_{ii}\right\rangle_{\mathcal{V}}$, noting that this quantity is proportional to the stored elastic energy. Where we evaluate the frequency spectra of flow statistics, these are denoted with a tilde $\widetilde{\left\langle\cdot\right\rangle}$. Where we report volumetric flow rates, these are normalised by the volumetric flow rate at $Wi=0$ for the same $Re$.

\subsection{Overview of dynamics}

\begin{figure}
\setlength{\unitlength}{1cm}
\begin{center}
\begin{picture}(7,6)(0,0)
 \put(0,0){\includegraphics[width=0.6\textwidth]{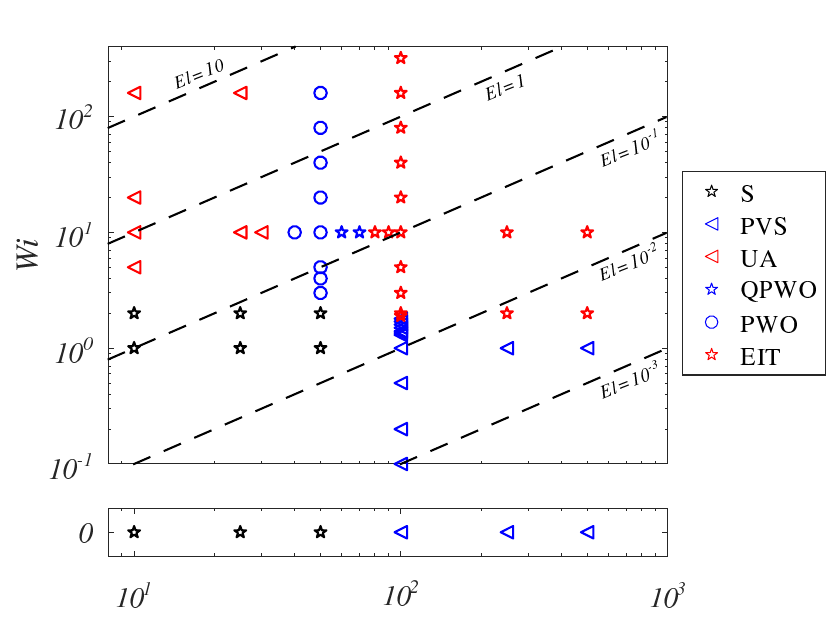}}
 \put(3.1,-0.0){{\emph{Re}}}  
\end{picture}
\end{center}
    \caption{Diagram of observed flow regimes in the $Re-Wi$ space: Black stars - Steady flow (S); blue triangles - Periodic vortex shedding (PVS); red triangles - Unsteady Arrowheads (UA); Blue circles - Periodic wake oscillations (PWO); blue stars - Quasi-periodic wake oscillations (QPWO); red stars - Elasto-inertial turbulence (EIT). Dashed black lines show isocontours of $El=Wi/Re$. \reva{Note the cluster of blue and red markers at $Re=100$ and $Wi\in\left[1,2\right]$ are all blue triangles (PVS) and red stars (EIT).}\label{fig:rewi}}   
\end{figure}

\begin{figure}
    \includegraphics[width=0.99\textwidth]{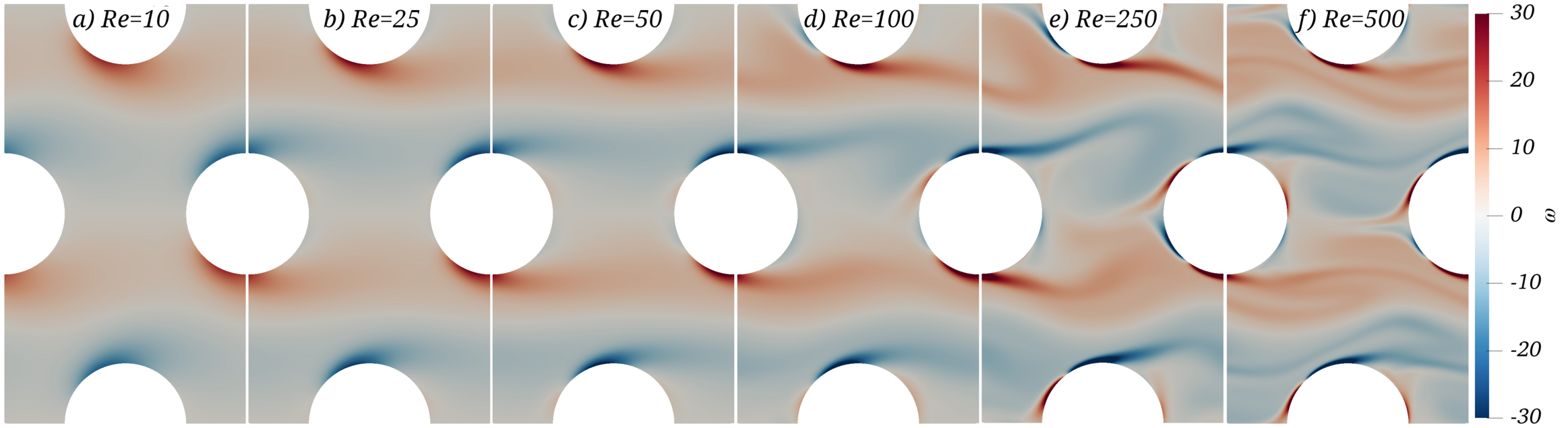}
    \includegraphics[width=0.99\textwidth]{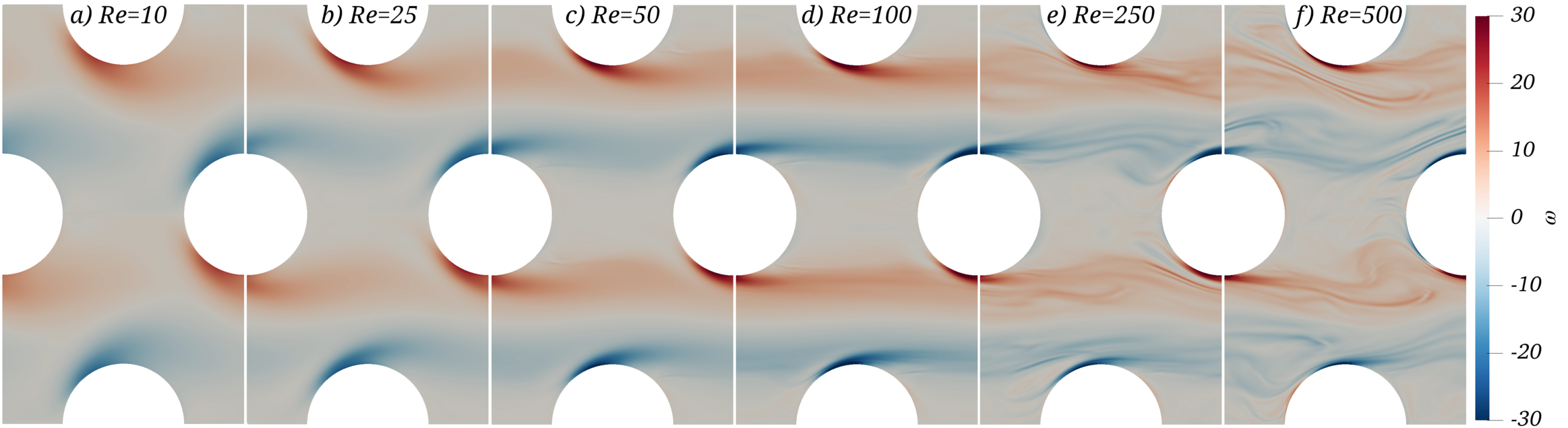}
    \caption{Instantaneous isocontours of vorticity $\omega$ for a range of $Re$ at $Wi=0$ (top panel), and $Wi=10$ (bottom panel).\label{fig:varRe_vort}}
\end{figure}

\begin{figure}
    \includegraphics[width=0.99\textwidth]{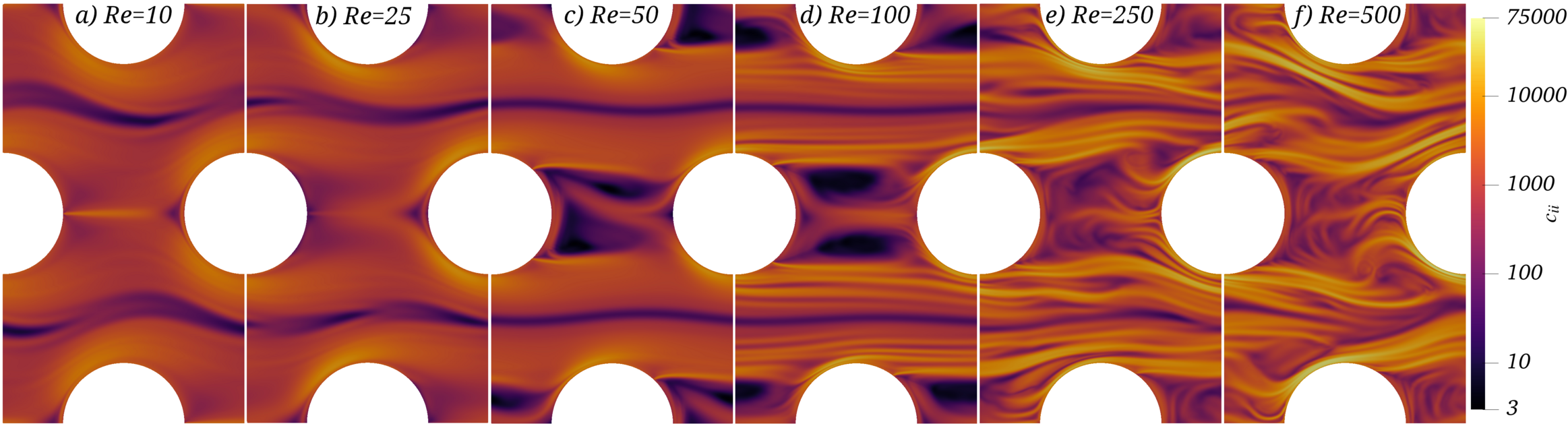}
    \caption{Instantaneous isocontours of conformation tensor trace $c_{ii}$ for a range of $Re$ at $Wi=10$.\label{fig:varRe_trc}}
\end{figure}

Figure~\ref{fig:rewi} shows the flow regimes found in our simulations across the $Re-Wi$ space. Figure~\ref{fig:varRe_vort} shows instantaneous isocontours of vorticity for a range of $Re$, in the Newtonian case (top panel) and at $Wi=10$ (lower panel). Figure~\ref{fig:varRe_trc} shows isocontours of conformation tensor trace $c_{ii}$ for $Wi=10$ and the same range of $Re$.

At small $Re\le50$ and $Wi\le2$ the flow is steady (black stars in figure~\ref{fig:rewi}, and panels a) to c) of the top row of figure~\ref{fig:varRe_vort}). In the Newtonian case, for $Re\ge100$ we see periodic vortex shedding (panels d) to f) of the top row of figure~\ref{fig:varRe_vort}). Whilst the simulations presented here are two-dimensional, we have run three-dimensional simulations with $Wi=0$ for the same range of $Re$. For $Re\le250$, the flow remains two-dimensional, whilst for $Re=500$ the flow is three-dimensional and turbulent. Thus we focus our investigation primarily on flows at $Re\le100$.

\begin{figure}
    \centering{\includegraphics[width=0.4\textwidth]{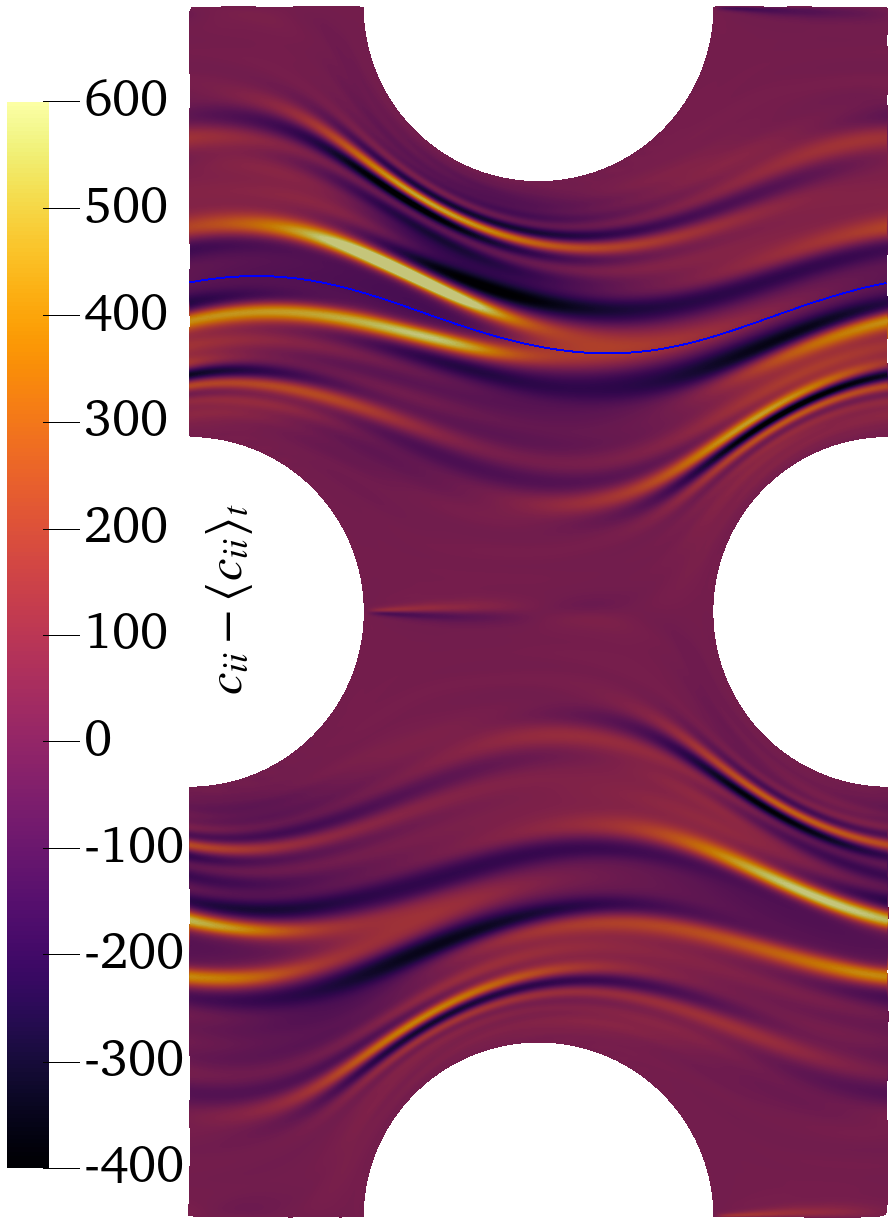}}
    \caption{\revb{Isocontours of conformation tensor trace $c_{ii}$ with the temporal mean subtracted, for $Re=10$ and $Wi=10$. The blue line indicates the Lagrangian pathline along which we calculate statistics.}\label{fig:arrowhead1}}
\end{figure}

For small $Re\le25$, as the Weissenberg number is increased we see arrowhead structures develop in the centre of the channels between the cylinders (panels a) and b) of figure~\ref{fig:varRe_trc})\revb{, and figure~\ref{fig:arrowhead1}, which shows isocontours of the instantaneous deviation from the temporal mean of $c_{ii}$, for $Re=10$ and $Wi=10$.}. As in plane Poiseuille flow~\cite{morozov_2022}, these arrowhead structures travel slightly slower than the velocity along the channel centreline. In this regime (red triangles in figure~\ref{fig:rewi}) the arrowhead structures interact: both through their finite length scale and the truncation of the domain by the imposition of periodicity, and by inducing fluctuations in the cylinder wakes, allowing the upper and lower channels to interact. Whilst we mostly observed the arrowheads in the upper and lower channels to be out of phase (as in panel a) in figure~\ref{fig:varRe_trc}), this is not always the case. Indeed, even in this small domain, multiple arrowhead structures can exist in a single channel. In the snapshot at $Re=25$ and $Wi=10$ (panel b) in figure~\ref{fig:varRe_trc}), there is a single arrowhead structure in the upper channel, and two arrowheads in the lower channel. While the motion of two out-of-phase arrowheads through the domain contributes to a signal in the spatially averaged statistics with a short time-scale - $T_{a}\approx{S}/2U_{max}\approx0.5$ - their interaction leads to phase shifts over much longer time-scales, of the order of tens of time units. We also note that in these minimal periodic unit domains, whilst the interaction of arrowheads may lead to quasi-periodic flow, we do not observe the development of chaotic flow or an ET-like state in our simulations. 

\begin{figure}
    \includegraphics[width=0.45\textwidth]{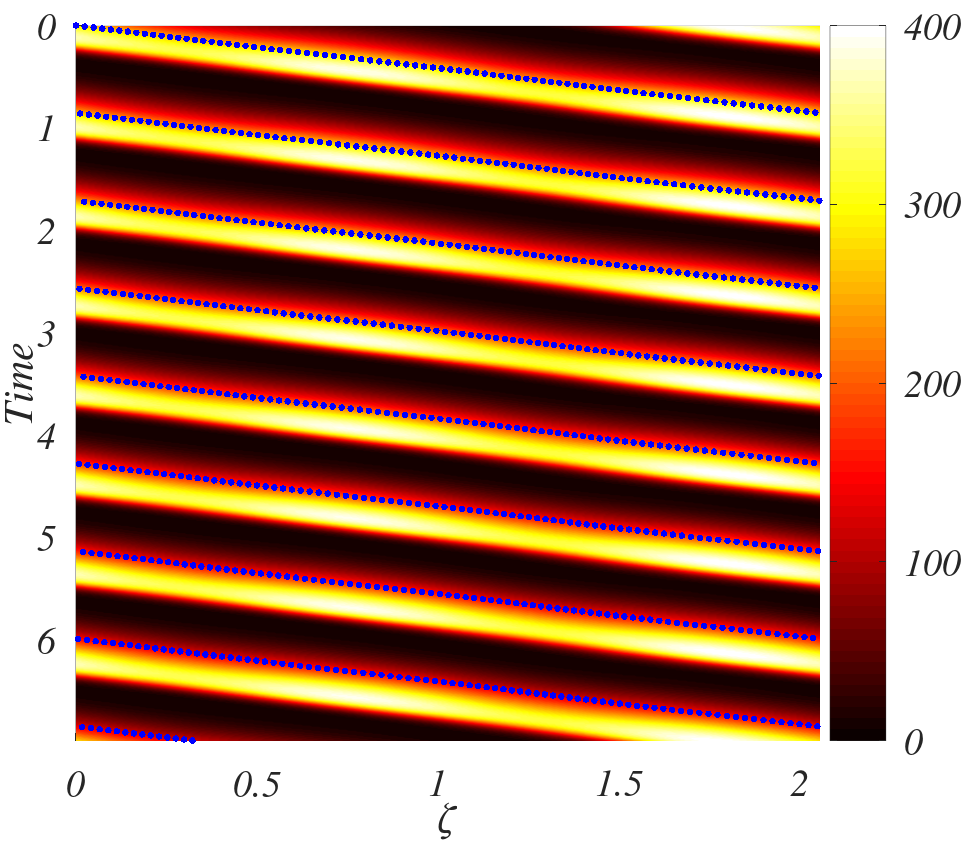}
    \includegraphics[width=0.54\textwidth]{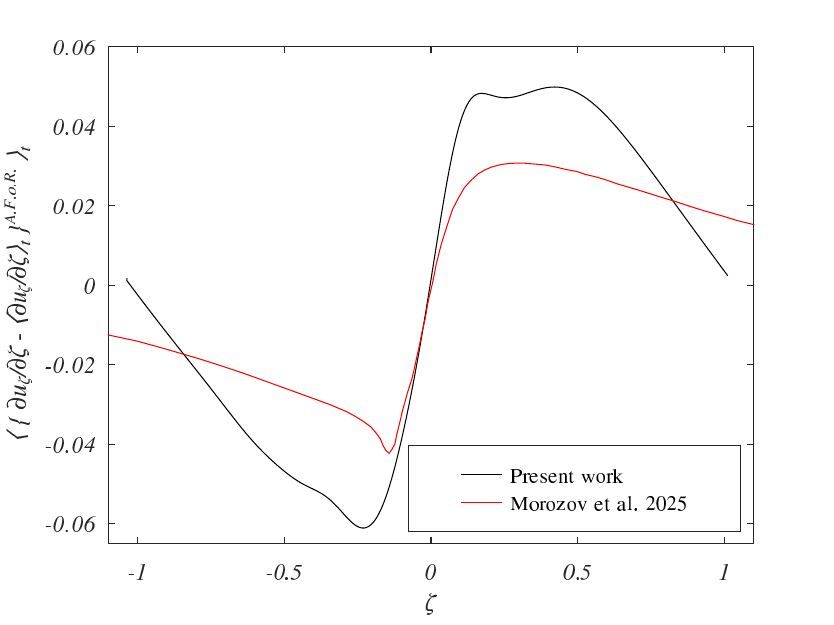}    
    \caption{\revb{Left panel: spatio-temporal evolution of $c_{ii}$ along a Lagrangian pathline in the centre of the upper channel. Blue dots represent the trajectory of a Lagrangian particle. Right panel: the average spatial distribution of the streamwise velocity gradient deviation in the frame of reference of the arrowhead. The red line shows data taken from figure 4c of~\cite{morozov_2025}.}\label{fig:arrowhead2}}
\end{figure}

%% New arrowheads section...
\revb{The Lagrangian pathline along the centre of the upper channel in shown in blue in figure~\ref{fig:arrowhead1}. We parameterise this pathline with the coordinate $\zeta$. The left panel of figure~\ref{fig:arrowhead2} shows the spatio-temporal evolution of $c_{ii}$ along this pathline, with the blue dotted line indicating the trajectory of a Lagrangian particle. The arrowhead is clearly a travelling wave structure embedded within the complex flow. The mean flow velocity along the pathline is $U_{0}=2.401$. In agreement with the results of~\cite{morozov_2022}, we observe that the arrowhead travels slightly slower than the mean flow velocity, with an average phase speed of $2.340=0.979U_{0}$. This is visible in the left panel of figure~\ref{fig:arrowhead2}, where the (dotted blue) Lagrangian trajectory and the spatio-temporal structure of $c_{ii}$ drift slightly over multiple passages across the domain. Along the pathline we denote the velocity magnitude as $u_{\zeta}$, and calculate $\partial{u}_{\zeta}/\partial\zeta$. At each time-instant, we then take the instantaneous deviation of this quantity from its temporal mean $\partial{u}_{\zeta}/\partial\zeta-\langle\partial{u}_{\zeta}/\partial\zeta\rangle_{t}$, and translate it to a frame of reference centered on the arrowhead
\begin{equation}\left\{\frac{\partial{u}_{\zeta}}{\partial\zeta}-\langle\frac{\partial{u}_{\zeta}}{\partial\zeta}\rangle_{t}\right\}^{\text{A.F.o.R.}},\end{equation}
in which the superscript $\text{A.F.o.R.}$ indicates the quantity has been shifted in $\zeta$ to a frame of reference moving with the arrowhead. We then take the temporal average of this phase-shifted quantity, to obtain the average spatial distribution (around the arrowhead) of the streamwise velocity gradient deviation. The resulting measure is equivalent to the measure shown in figure 4c of~\cite{morozov_2025} for arrowheads in a channel flow. The right panel of figure~\ref{fig:arrowhead2} shows this quantity (in black) alongside the data from figure 4c of~\cite{morozov_2025} (in red). Note, the data from~\cite{morozov_2025} have been scaled to account for the choice of non-dimensionalisation. In both cases we see the same overall structure, with two local stagnation-points, a similar peak magnitude and decay away from the arrowhead (although note that the streamwise extent of our domain is shorter than that used in~\cite{morozov_2025}.) Detailed differences in the profiles of $\partial{u}_{\zeta}/\partial{\zeta}$ may be attributed to geometrical effects: in the present work, we have arrowheads travelling along a curved streamline, in an effective channel with non-uniform cross section.}

\begin{figure}
\begin{center}   
\includegraphics[width=0.8\textwidth]{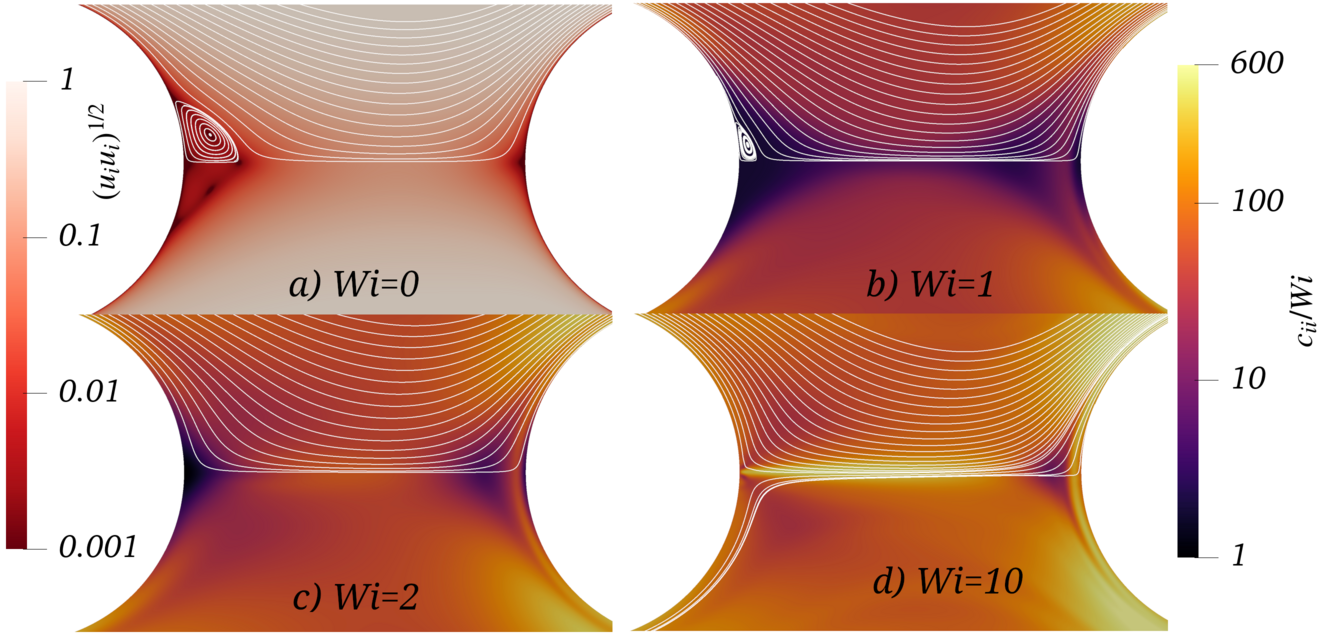}
\caption{Flow snapshots at $Re=10$ for various $Wi$. Panel a) shows streamlines, and isocontours of velocity magnitude at $Wi=0$. Panels b) to d) show streamlines and isocontours of conformation tensor trace (normalised by $Wi$) for b) $Wi=1$, c) $Wi=2$, and d) $Wi=10$.\label{fig:re10_snap}}
\end{center}
\end{figure}

At low $Re$, elasticity suppresses separation in the cylinder wake. Figure~\ref{fig:re10_snap} shows flow streamlines for $Re=10$ and a range of $Wi$ in the region between two cylinders. At $Wi=0$, a clear \revb{inertia-driven} separation bubble exists in the cylinder wake. At $Wi=1$ this separated region has reduced in size, and for $Wi\ge2$ there is no separation. Note that at $Wi=10$ the flow has broken symmetry, with streamlines crossing the line between the two cylinder centres. \revb{This symmetry breaking in the wake between two cylinders has been observed in previous numerical and experimental studies of viscoelastic flow between two cylinders in the creeping flow regime (see \citealt{varshney_2017, kumar_2021b}), however at $Re=10$ we do not observe the development of separation bubble arising due to elastic effects even at $Wi\geq 160$.}

At $Re=50$, the separated wake region extends the full distance between cylinders, and consists of two counter-rotating vortices (panels c) of figures~\ref{fig:varRe_vort} and~\ref{fig:varRe_trc}). For $Wi\le2$ these are steady, whilst at higher $Wi$ these show periodic oscillations in size and location (indicated by blue circles in figure~\ref{fig:rewi}). At $Re=50$, we do not observe arrowhead structures at any $Wi$ simulated. This behaviour is discussed further in~\S~\ref{sec:re50}.

At $Re=100$ and $Wi\le1.8$ we observe vortex shedding where the recirculating vortices interact with the next cylinder downstream. At higher $Wi$, a chaotic flow state is observed, with the hallmarks of elasto-inertial turbulence (EIT)\revb{: it requires both a critical amount of elasticity and inertia, and contains a spanwise structure reminiscent of EIT in channels}. This EIT state is observed at $Re=100$ for all simulated $Wi\in\left[1.9,320\right]$, and for all simulated $Re>100$, $Wi\ge2$ (illustrated by the red stars in figure~\ref{fig:rewi}). We note here that for the present geometry, the parameter space in which we observed arrowheads (red triangles in figure~\ref{fig:rewi}) is not connected to the region where we observe EIT (red stars in figure~\ref{fig:rewi}). We discuss this further in~\S~\ref{sec:wi10} and~\S~\ref{sec:re50}. \revc{In addition, we have run $3D$ simulations starting from the two-dimensional EIT states (at parameter values $Re=100, Wi=10$), with the third dimension being $4$ cylinder diameters in extent. Spanwise structures did not grow, with the maximal magnitude of spanwise velocity remaining below the residual error floor of the discretisation scheme. In the rest of the manuscript we therefore focus on $2D$ simulations, in a similar manner to \cite{kumar_2023}.}

\subsection{Transition pathway to EIT with increasing elasticity}

\begin{figure}
\setlength{\unitlength}{1cm}
\begin{center}
\begin{picture}(14,5)(0,0)
 \put(0,0){\includegraphics[width=0.49\textwidth]{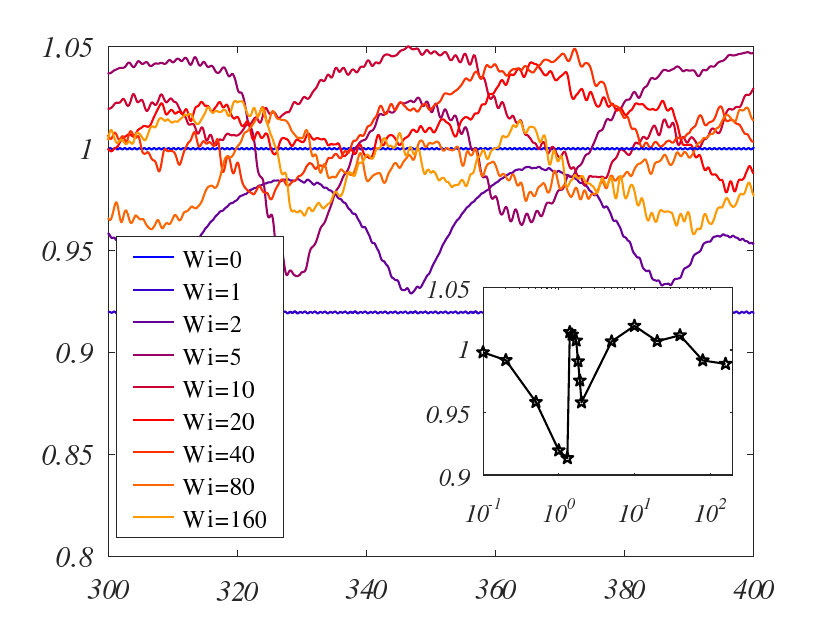}}
 \put(7,0){\includegraphics[width=0.49\textwidth]{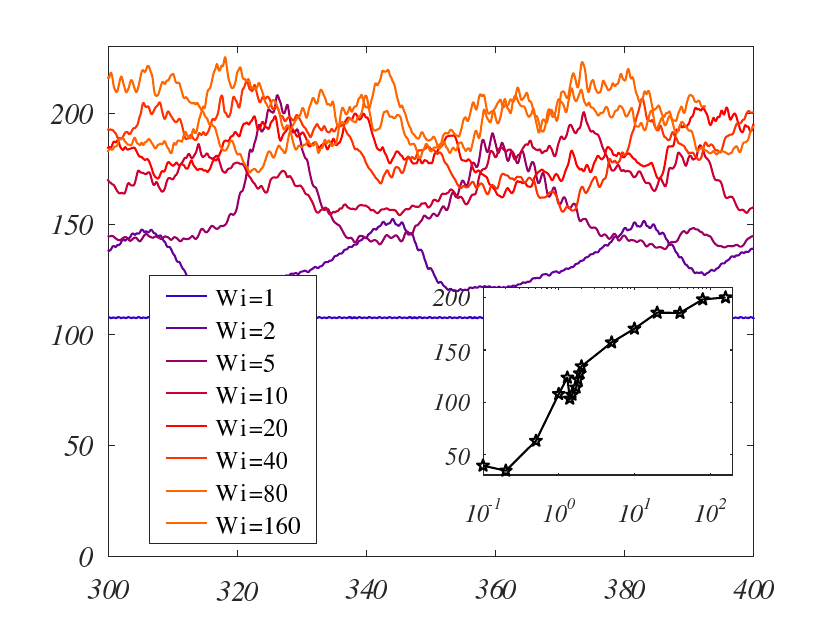}}
 \put(0.0,2.5){\rotatebox{90}{\scriptsize{$\dot{Q}$}}}
 \put(7,2.2){\rotatebox{90}{\scriptsize{$E_{E}/Wi$}}}    
 \put(11.45,1.4){\tiny{$\left\langle{E}_{E}/Wi\right\rangle_{t}$}}
 \put(5.0,1.4){\tiny{$\left\langle\dot{Q}\right\rangle_{t}$}}
 \put(3.0,0){\scriptsize{\emph{Time}}} 
 \put(10.0,0){\scriptsize{\emph{Time}}} 
 \put(4.55,0.7){\tiny{\emph{Wi}}}
 \put(11.55,0.7){\tiny{\emph{Wi}}}     
\end{picture}
\end{center}
\caption{Time evolution of normalised volumetric flow rate $\dot{Q}$ (left) and $E_{E}/Wi$ (right) for $Re=100$ and a range of $Wi$. The insets show the variation with $Wi$ of the temporal averages of these quantities. Note more values of $Wi$ are used for the inset than are plotted in the main figures.\label{fig:re100_t}}
\end{figure}

\begin{figure}
\setlength{\unitlength}{1cm}
\begin{center}
\begin{picture}(14,5)(0,0)
 \put(3.5,0){\includegraphics[width=0.49\textwidth]{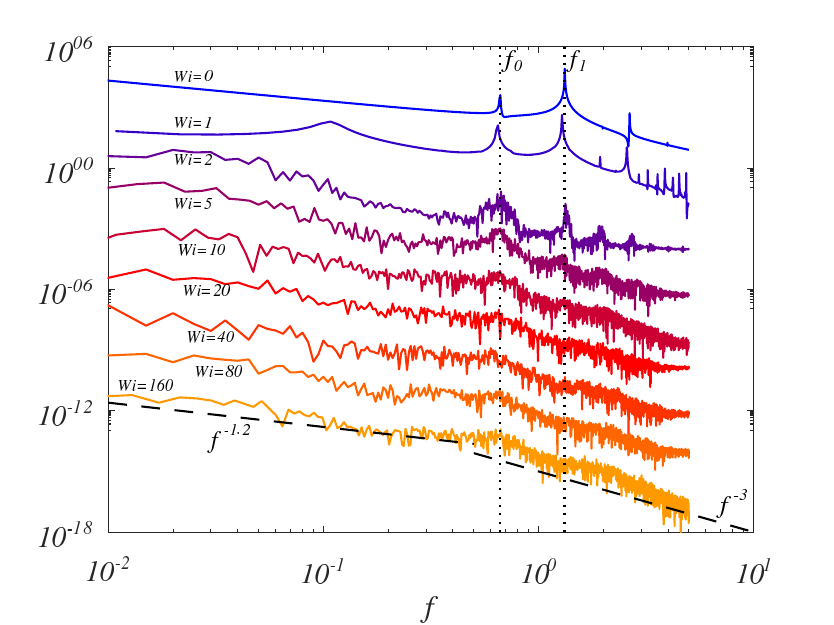}}
 \put(3.5,2.5){\rotatebox{90}{\scriptsize{$\widetilde{E_{K}}$}}}    
\end{picture}
\end{center}
\caption{Frequency spectrum of the spatially averaged kinetic energy $\widetilde{E_{K}}$ for $Re=100$ and a range of $Wi$. Note the lines have been vertically shifted to allow for ease of visualisation. \reva{The dotted vertical lines indicate the vortex shedding frequency $f_{0}$ and its first harmonic $f_{1}$.}\label{fig:re100_f}}
\end{figure}

We next focus on the transition pathways to EIT. We fix $Re=100$ and vary $Wi$. The left panel of figure~\ref{fig:re100_t} shows the time evolution of the volumetric flow rate $\dot{Q}$ (normalised by the mean volumetric flow rate at $Wi=0$). The right panel of figure~\ref{fig:re100_t} shows the time evolution of the volume-averaged conformation tensor trace $E_{E}$ for a range of $Wi$. Figure~\ref{fig:re100_f} shows the frequency spectrum $\widetilde{E_{K}}$ of the volume averaged kinetic energy $E_{K}=\left\langle\rho{u}_{i}u_{i}\right\rangle_{V}/2$.

For $Wi=0$ and $Wi=1$ the flow rate $\dot{Q}$ is almost steady, with only small periodic oscillations due to the vortex shedding. The same behaviour is seen in the volume-averaged conformation tensor trace $E_{E}$ at $Wi=1$, and can be seen in the top two lines of the frequency spectra in figure~\ref{fig:re100_f}, with peaks at two frequencies at $f_{0}\approx0.65$ and $f_{1}\approx1.3$. There is a significant decrease in flow rate $\dot{Q}$ from $Wi=0$ to $Wi=1$. Although the flows at these low $Wi$ are qualitatively similar, elasticity results in an increase in flow resistance. At $Wi=2$ the flow is chaotic, although the dynamics remains dominated by the vortex shedding (short time-scale motion), and longer time-scale (approx $40$ time units) oscillations, which can be seen in the frequency spectra in figure~\ref{fig:re100_f}. For larger $Wi$, a greater range of frequencies are present: the spectra show a power law, with slope approximately $f^{-1.2}$ at frequencies below the vortex-shedding frequency $f_{0}$, and of approximately $f^{-3}$ at higher frequencies. The exponent of $-3$ \reva{matches that of the viscoelastic jet studied by}~\citep{giulio_2026}. \reva{We note that~\cite{giulio_2026} showed that the temporal power spectra in elastic turbulence is not universal, but flow dependent, and it is possible that the same is true of elasto-inertial turbulence.} The insets of figure~\ref{fig:re100_t} show the temporal averages in $\dot{Q}$ and $E_{E}/Wi$ as $Wi$ is varied. Below the transition to EIT, there is a clear discontinuity in both quantities. At approximately $Wi=1.4$, there is a $10\%$ increase in $\dot{Q}$, and a corresponding decrease in $E_{E}/Wi$, indicative of some kind of transition in the dynamics to a higher kinetic energy, lower elastic energy state. At higher $Wi$, $\dot{Q}$ decreases (and $E_{E}/Wi$ correspondingly increases) continuously with increasing $Wi$, up to the transition to EIT at approximately $Wi=2$.

\begin{figure}
\setlength{\unitlength}{1cm}
\begin{center}
\begin{picture}(14,5.3)(0,0)
\put(0,0){\includegraphics[width=0.505\textwidth]{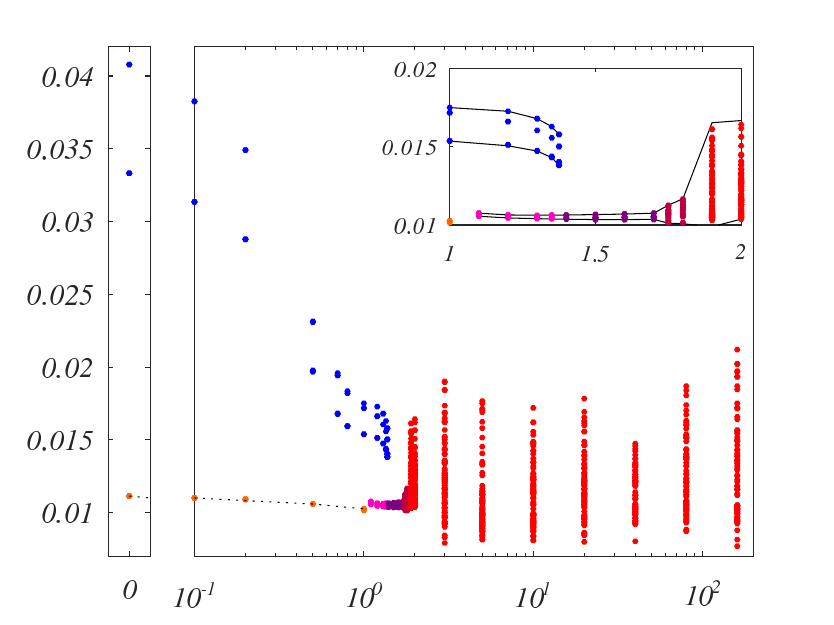}}
\put(7,0){\includegraphics[width=0.49\textwidth]{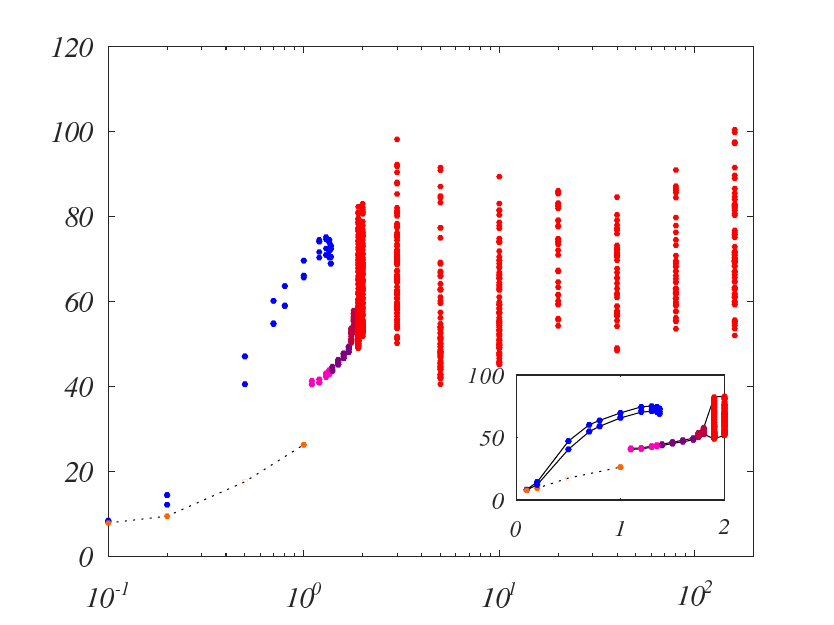}}
 \put(-0.1,2.3){\rotatebox{90}{\scriptsize{$\mathcal{A}_{u}$}}}
 \put(7,2.2){\rotatebox{90}{\scriptsize{$\mathcal{A}_{c}$}}}    
 \put(3.6,0){\scriptsize{$Wi$}} 
 \put(10.0,0){\scriptsize{$Wi$}} 
\end{picture}
\end{center}
    \caption{Bifurcation diagrams for $Re=100$ with varying $Wi$, for two measures: $\mathcal{A}_{u}=\left\langle{u}_{2}u_{2}\right\rangle_{\mathcal{V}}^{1/2}$ (left panel) and $\mathcal{A}_{c}=\left\langle{c}_{12}^{2}\right\rangle_{\mathcal{V}}^{1/2}/Wi$ (right panel). For each value of $Wi$, the maxima and minima of $\mathcal{A}_{u}$ and $\mathcal{A}_{c}$ are plotted.\label{fig:re100_bifurc}}
\end{figure}

To explore the transition pathway in more detail, we construct a bifurcation diagram, which is shown in figure~\ref{fig:re100_bifurc}. We plot the \revb{local} maxima and minima of $\mathcal{A}_{u}=\left\langle{u}_{2}u_{2}\right\rangle_{\mathcal{V}}^{1/2}$ (left panel) and $\mathcal{A}_{c}=\left\langle{c}_{12}^{2}\right\rangle_{\mathcal{V}}^{1/2}/Wi$ (right panel) for each value of $Wi$ simulated. \reva{Note that a given periodic flow can have multiple local maxima/minima of $\mathcal{A}_u,\mathcal{A}_c$ due to the presence of subharmonic modes.} \reva{These metrics are chosen as proxies for inertial and elastic effects and the identified transition pathway is independent of the choice of metric.}

At $Wi=0$ there is an unstable branch (orange circles, connected by a dashed line) corresponding to steady laminar flow with no vortex shedding, and a stable branch (blue dots), with larger $\mathcal{A}_{u}$ in which there is periodic vortex shedding. We are only able to observe the unstable branch when starting simulations from zero flow, \revb{and we do not remain on this branch indefinitely:} we see deviations from this branch grow exponentially, with a growth rate which increases with increasing $Wi$. We are unable to track this unstable branch beyond $Wi=1$.

For the stable upper branch, as $Wi$ is increased $\mathcal{A}_{u}$ decreases, and $\mathcal{A}_{c}$ increases. This reduction in the component of the kinetic energy aligned perpendicular to the flow direction corresponds to the decrease in $\left\langle\dot{Q}\right\rangle_{t}$ and increase in $\left\langle{E}_{E}/Wi\right\rangle_{t}$ seen in the insets of figure~\ref{fig:re100_t}. As $Wi$ is increased along this stable branch, the kinetic energy of the flow decreases, and the elastic energy stored in the flow increases.

As $Wi$ is increased this branch disappears via a sub-critical saddle-node bifurcation at $Wi=1.4$, and there is a transition to a lower stable branch (purple circles in figure~\ref{fig:re100_bifurc}), in which the volumetric flow rate is larger (see the jump in $\left\langle\dot{Q}\right\rangle_{t}$ in the inset of the left panel of figure~\ref{fig:re100_t}), and the transverse oscillations in the wake of the cylinder are reduced. We are able to trace this stable lower branch branch smaller $Wi$ \reva{(pink circles in figure~\ref{fig:re100_bifurc})} as far as $Wi=1.1$ \revc{implying that there are two distinct stable branches for $1.1\leq Wi\leq 1.4$}. Efforts to trace this branch to lower $Wi$ were unsuccessful, and resulted in a transition back to the upper branch. \revb{The flow states of both upper and lower branches are dominated by vortex shedding dynamics.}

\begin{figure}
\begin{center}
\includegraphics[width=0.32\textwidth]{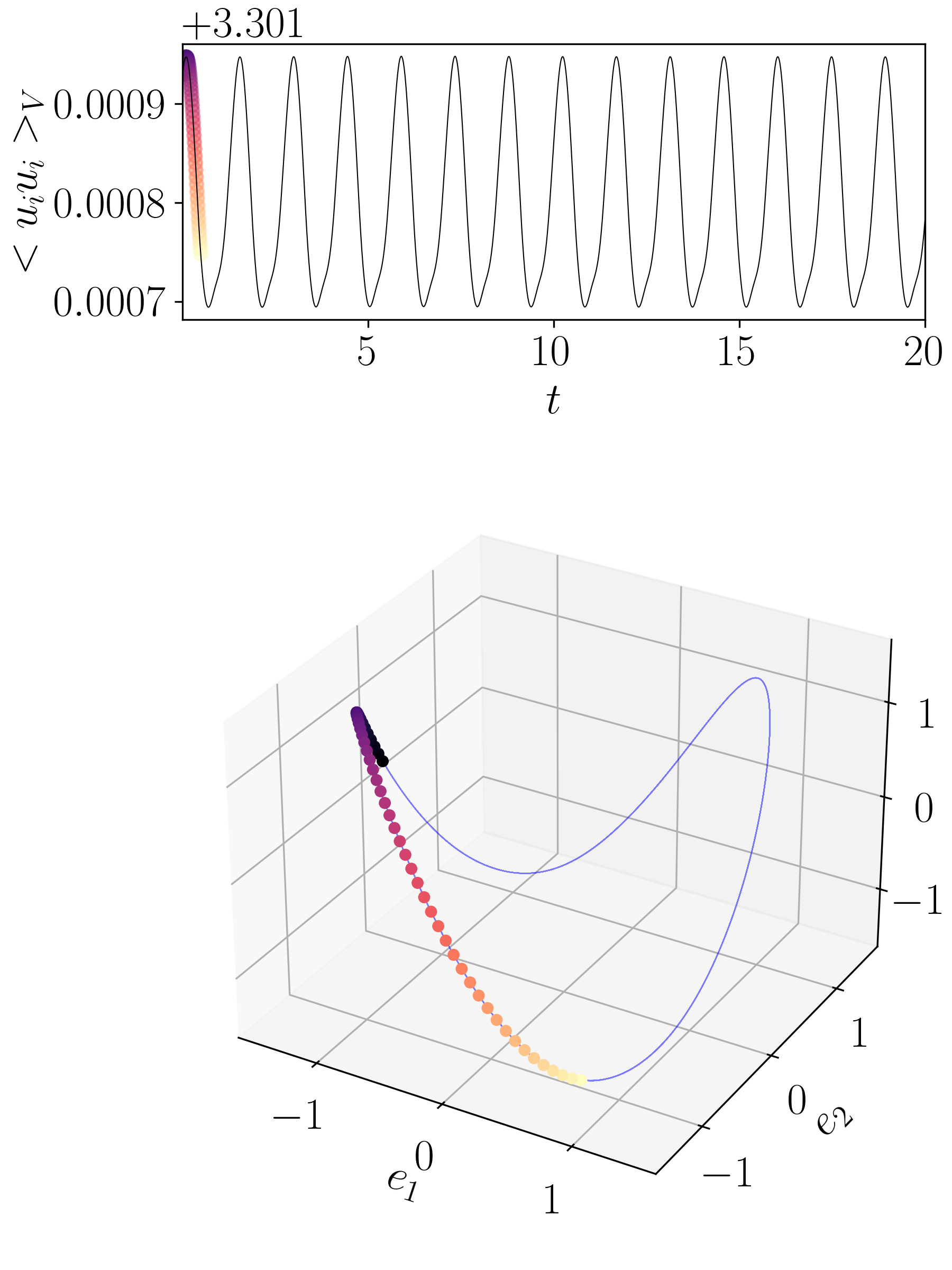}
\includegraphics[width=0.32\textwidth]{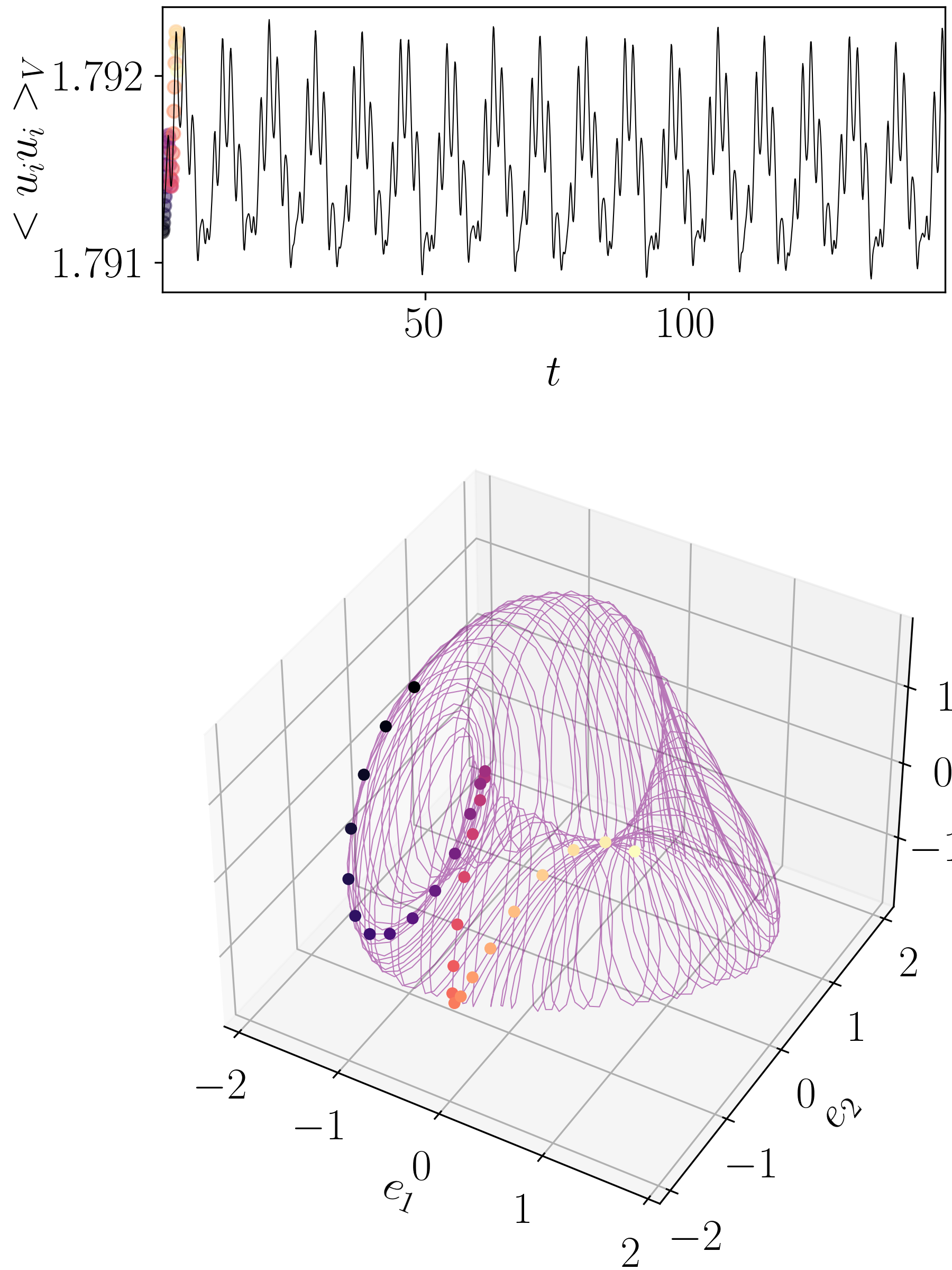}
\includegraphics[width=0.32\textwidth]{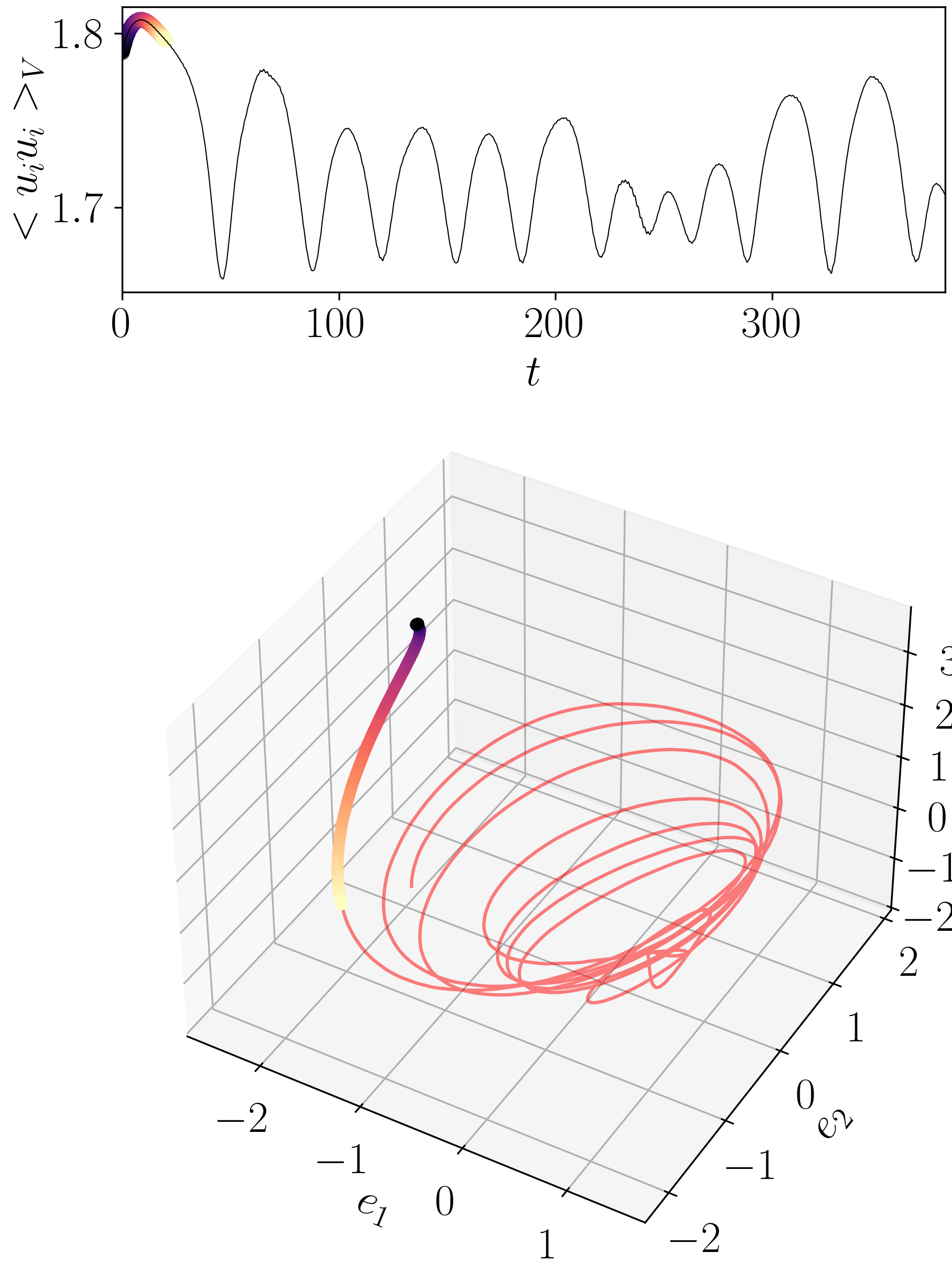}
\end{center}
\caption{Approximation of the attractors for the lower branch PO at $Wi=1.35$ (left), the $\mathbb{T}^2$-torus at $Wi=1.75$ (middle), and the chaotic solutions at $Wi=2$ (right). Top: time-series of the \textit{rms} of the volume-integrated kinetic energy. Bottom: attractors approximated by 20 embeddings of the time series above. The coordinates are aligned with the principal coordinates of the attractor and rescaled for illustrative purposes. Dots illustrate the direction of time on the attractor. The attractors illustrate the Ruelle-Takens-Newhouse route to chaos.  \label{fig:embeddings}}
\end{figure}

As $Wi$ is further increased from this lower branch solution a Ruelle-Takens-Newhouse (RTN) route to chaos is observed, supported by a series of supercritical bifurcations~\citep{ruelle_1971,Drazin_2002}, with the flow transitioning from a periodic orbit (lower branch solution) at $Wi=1.35$, to a torus at $Wi=1.75$, and chaos for $Wi\ge1.9$. This is further evidenced through the absence of hysteresis in EIT; we were unable to continue chaotic solutions by time-stepping starting from a $Wi>1.9$ solution and moving down in $Wi$. We generate an approximation of the geometry of the visited attractors on the path to EIT via time embeddings~\citep{takens2006detecting,broomhead1986extracting} of the signal of the volume-integrated \textit{rms} values of the kinetic energy. The time series used to generate the embeddings and approximate the attractors are shown in figure~\ref{fig:embeddings} (top). Taken's embeddings theorem guarantees that $m>2d+1$ embeddings is sufficient to avoid self-intersections of the trajectories. We choose $m=20$ embeddings which proved sufficient to represent our attractors, and a time offset $\tau$ determined by the first zero crossing of the time signal autocorrelation (i.e. the signal is uncorrelated with itself), except for the turbulent case, where the offset is set to $\tau=2.1$, and we just aim to illustrate the aperiodic behaviour. The attractor coordinates $\{e_i\},\ i=1,2,3$, are then obtained following~\cite{broomhead1986extracting}, where we perform a singular value decomposition (SVD) of the embeddings matrix, and project our embeddings onto the principal axes of the attractor. We rescale each coordinate by its standard deviation for visualisation purposes. The captured attractors are shown in figure~\ref{fig:embeddings} (bottom), and we observe that the lower branch solution at $Re=100$, $Wi=1.35$ corresponds to a simple periodic orbit, which then bifurcates to a $\mathbb{T}^2$-torus by $Wi=1.75$ and finally becomes aperiodic at $Wi=2$ confirming the RTN route to chaos.

One can consider the flow in this geometry as two channels, confined by cylinders and the cylinder wakes. The vortex shedding behind the cylinders provides a sufficient perturbation to the flow in the channels to trigger EIT. The flow in the channels and the recirculating wakes then interact. The power law of the energy spectra with slope $-3$ above the vortex shedding frequency $f_{0}$ is driven by the EIT-like flow in the channels. The velocity magnitudes within the recirculating wake are much smaller, and there the dynamics occur at longer time-scales, leading to a different slope in the energy spectra ($-1.2$ for $f<f_{0}$). This view is supported by our simulations are $Re=50$ (in~\S~\ref{sec:wi10}), where there are wake oscillations but no vortex shedding, and the motion occurs over long timescales. \revc{We note that these dynamics are different to those observed by \cite{zhu_2024}, however the geometries in the two are distinctly different and the arrowhead structures observed by \cite{zhu_2024} are artificially triggered, not naturally evolving like those considered herein. Furthermore, they use a different constitutive model, and impose different boundary conditions on the cylinder walls.}

\subsection{Koopman analysis}
\reva{To identify what flow dynamics can be associated with transition to EIT, we perform a Koopman analysis on our simulations.} Koopman analysis was introduced by~\cite{koopman1931hamiltonian,koopman1932dynamical} (and recently reviewed in~\cite{mezic2013analysis}). It is a powerful technique to study nonlinear systems by lifting the state-vector to a set of observables which are evolved under an infinite-dimensional linear operator. We apply this technique to gain further insight into the self-sustaining mechanisms observed in our simulations and the spatio-temporal structures on the solutions obtained en route to EIT. 

We consider a dynamical system in \eqref{eq:mass}-\eqref{eq:cte}, which is observed on discrete time steps according to a function $F:\Gamma\to \Gamma$,
\begin{equation}
q_{n+1} = F(q_n), \quad n\geq 0,    
\end{equation}
from an initial state $q_0$, \revb{where $\Gamma$} is an appropriate domain. The Koopman operator for a system $\mathcal{K}$, is defined by its action on the observed system, $g: \Gamma \to \mathbb{R}$, such that
\begin{equation}
    \mathcal{K}g = g\circ  F, 
\end{equation}
and in turn,
\begin{equation}
    \mathcal{K}g(q_n)= g(F(q_n)) = g(q_{n+1}), \label{eq:koopman}
\end{equation}
advances the system one step forward in the observed time. Note that this formalism is general to a suitable set of observables or functions as detailed in~\cite{williams2015data, schmid2022dynamic, colbrook2023residual}. As shown in~\eqref{eq:koopman} Koopman analysis provides a convenient description of nonlinear systems as an infinite-dimensional linear operator. The spectral properties of the Koopman operator are key to the linearisation process, however, an analytical treatment of the Koopman eigenfunctions is rarely available~\citep{bagheri2013koopman}, and we have to resort to numerical approximations to compute the spectral properties of $\mathcal{K}$. Dynamic mode decomposition (DMD)~\citep{schmid2010dynamic} provides a way to approximate the eigenvalues and eigenvectors of the Koopman operator directly from data (see~\cite{schmid2022dynamic} for a detailed discussion), and has been extended and improved since its inception. In this work we use the algorithms described in the recently developed ResDMD~\citep{colbrook2023residual}, which compute spectra and pseudo spectra of general Koopman operators with error control. \revb{We pay particular attention to choice of observables so that they are appropriate for viscoelastic flows. Following \cite{kumar2025elastoinertial}, we consider $\bm{q}=[\bm{u},\bm{T}]$, where $T_{ij}=\sqrt{(1-\beta)/(\textit{Re}\textit{Wi})}\theta_{ij}$ is the weighted symmetric square root of the conformation tensor (with $\theta_{ik} \cdot \theta_{kj}=c_{ij}$), also called the stretch tensor. We consider $\bm{q}$ as an appropriate state variable so that 
\begin{equation}
    \langle q,q\rangle = 2U_{tot} = 2(U_K +U_E) = \int_\Omega \left( u_iu_i + \frac{1-\beta}{\textit{Re}\textit{Wi}} \theta_{kj}\theta_{kj} \right),
\end{equation}
where the term $\theta_{kj}\theta_{kj}$ represents the exact mechanical energy for Hookean dumbbells \cite{wang2014time} and is here inherited even with a finite $\varepsilon$, and $\Omega$ denotes the physical domain. We choose to omit the density from the state vector even though the system is weakly compressible due to the computational cost of the analysis. No differences were observed in the subsequent results.

We then collect data for a sufficiently long single ergodic trajectory. We view the trajectory data as,
\begin{align}
B = \left(\begin{matrix}
\sqrt{W_\alpha
}q_0 & \sqrt{W_\alpha
}q_1 & \dots & \sqrt{W_\alpha
}q_{M+1}
\end{matrix}
\right),    
\end{align}
where $\sqrt{W_{\alpha}}$ represent the spatial quadrature on the node distribution in our non-uniform grid and each flow field is flattened to a vector of dimension $q\in\mathbb{R}^{N_q}$, with $N_q=N_n\times 5$, and $M$ denotes the number of observations. We then denote $\bm{X}=({b_0 \dots b_{M}})$ and $\bm{Y}=({b_1 \dots b_{M+1}})$ the enumerations of the first $M$ columns, and the second to final columns respectively, such that
\begin{equation}
    \bm{Y}=F(\bm{X}).
\end{equation}

We consider our DMD observables $\Psi_{\text{DMD}}= \bm{X}$. Let }

\begin{equation}
    \sqrt{W}\Psi_{\text{DMD}}^T \approx U_r\Sigma_r V_r^T,
\end{equation}
where $\Sigma_r\in \mathbb{R}^{r\times r}$ is a strictly positive diagonal matrix, and $V_r\in \mathbb{R}^{N_q\times r}$, $U_r\in \mathbb{R}^{(M+1)\times r}$ satisfy that $V_r^TV_r=U_r^TU_r=I_r$, where $r$ is the truncation of choice, \revb{amounting to a projection onto the leading VEPOD (viscoelastic proper orthogonal decomposition) modes \citep{kumar2025elastoinertial,holmes2012turbulence}}. $W$ denotes the integration weights in~\cite{colbrook2023residual} which have components $w^{(i)}=1/M$ for an ergodic sampling. The transpose of $\Psi_{\text{DMD}}$ is considered so that ResDMD can be fitted within a Galerkin framework~\citep{colbrook2023residual}. We then form the matrices
\begin{align}
\Psi_X &= X^T V_r \Sigma_r^{\dagger}, \\
\Psi_Y &= Y^T V_r \Sigma_r^{\dagger},
\end{align}
where $\dagger$ denotes a pseudoinverse. These matrices can now be directly used in algorithms 1 and 2 in~\cite{colbrook2023residual} to compute the approximation of the Koopman eigenvalues and eigenvectors $(\bm{g},\lambda)$ as well as upper bounds on their residuals. Although we acknowledge the problem of vanishing residuals in ResDMD, we intentionally keep $r\ll M$ so that residuals can be properly approximated and we only require one set of snapshots.

\begin{figure}
\setlength{\unitlength}{1cm}
\begin{center}
\begin{picture}(14,3)(0,0)
 \put(0,0){\includegraphics[width=0.32\textwidth]{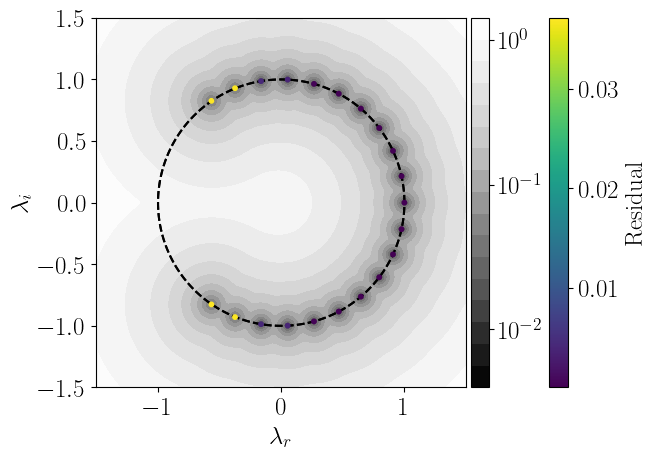}}
\put(4.666667,0){\includegraphics[width=0.32\textwidth]{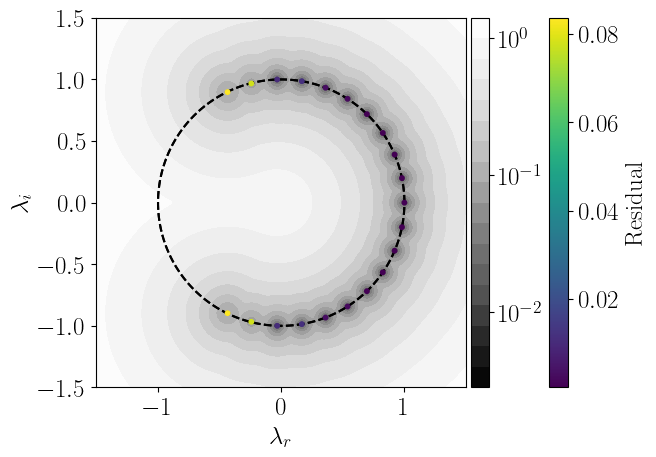}}
\put(9.333333,0){\includegraphics[width=0.32\textwidth]{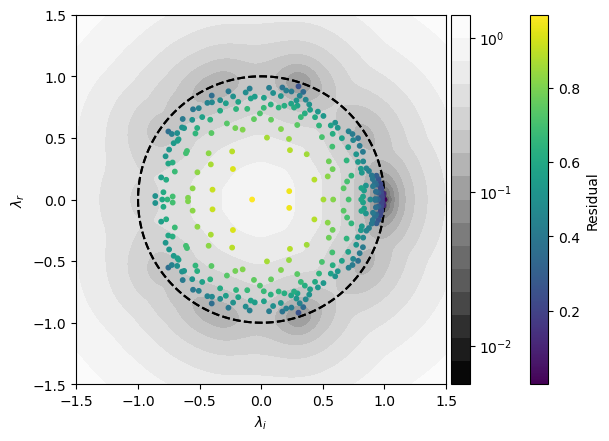}}
\put(9.27, 1.5){\rotatebox{90}{\colorbox{white}{\tiny{$\lambda_i$}}}}
\put(11, -0.03){{\colorbox{white}{\tiny{$\lambda_r$}}}}
\put(13.45, 1.1){\rotatebox{90}{\colorbox{white}{\tiny{Residual}}}}
\end{picture}
\end{center}
\caption{Pseudospectra contours for the residuals of the approximation of the Koopman eigenvalues and computed eigenvalues with residuals (colour). Left: Lower branch periodic orbit at $Wi=1.35$, Middle: Upper branch periodic orbit at $Wi=1.35$. Right: EIT at $Wi=10$, $Re=100$ in all. \label{fig:koopman_evals}}
\end{figure}

\begin{figure}
\begin{center}
\includegraphics[width=0.75\textwidth]{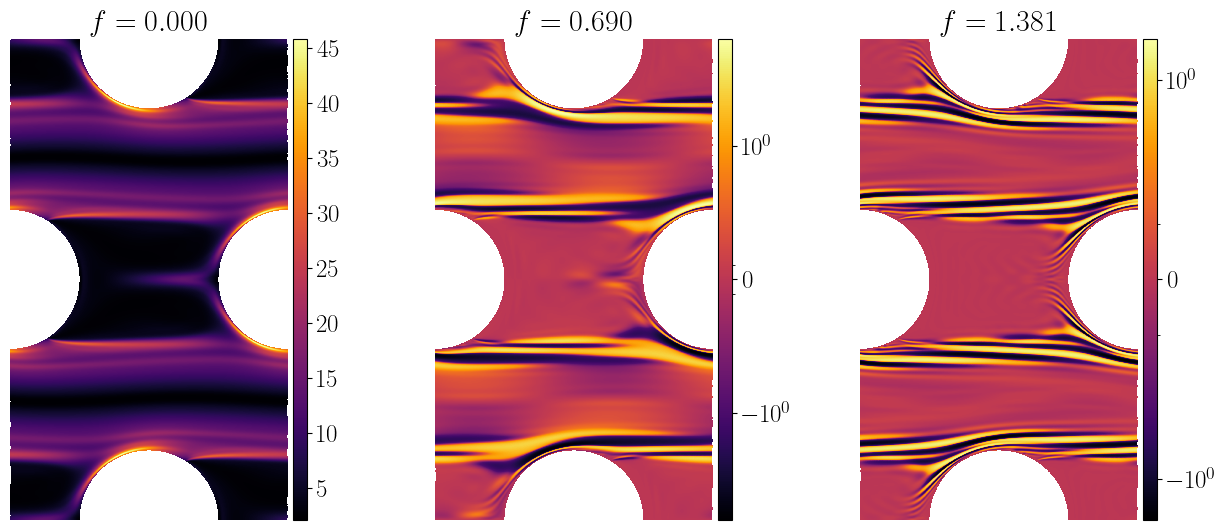}\\
\includegraphics[width=0.75\textwidth]{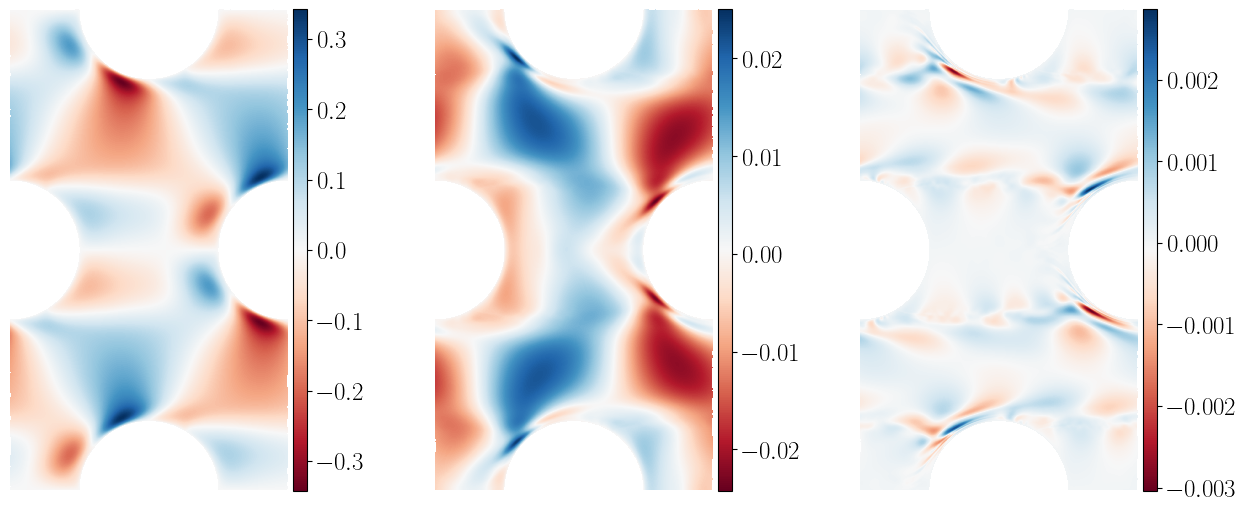}
\end{center}
\caption{Contours of the Koopman modes from the lower branch periodic orbit at $Re=100$, $Wi=1.35$. Top: $c_{ii}$, symmetric logarithmic colour scale has been used except for a small region around 0. Bottom: $u_{2}$. Left: Mean flow, middle: real part of the fundamental oscillatory frequency, right: first harmonic.\label{fig:koopman_PO_emodes_lb}}
\end{figure}

\begin{figure}
\begin{center}
\includegraphics[width=0.75\textwidth]{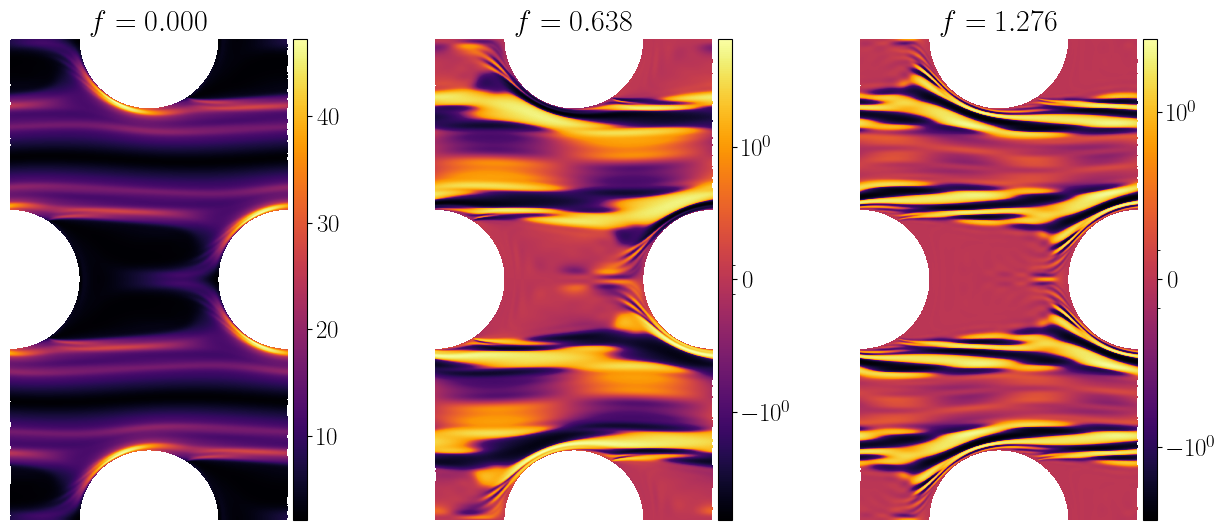}\\
\includegraphics[width=0.75\textwidth]{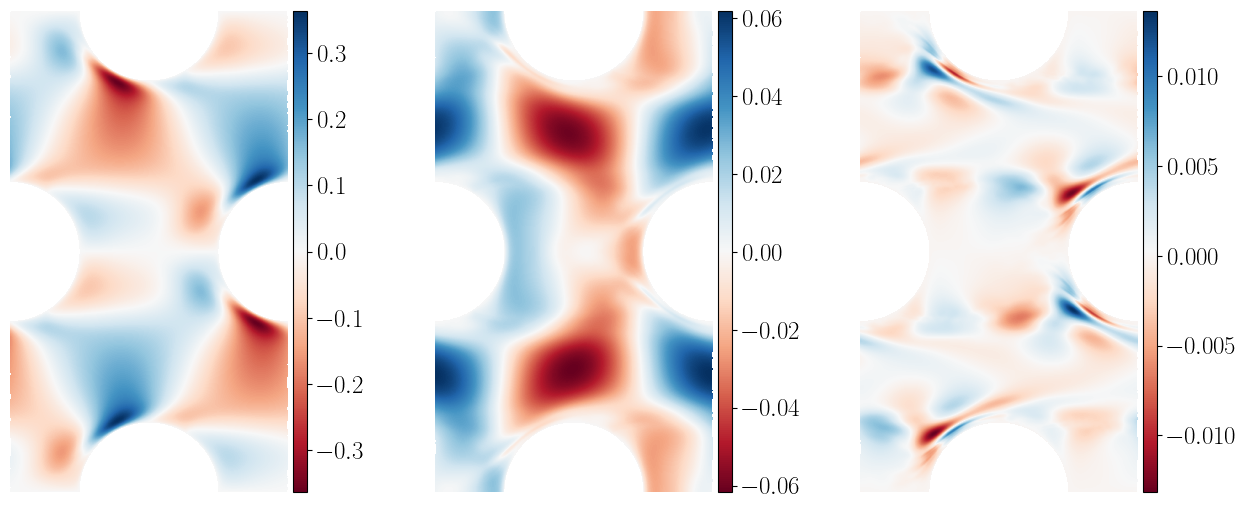}
\end{center}
\caption{Contours of the Koopman modes from the upper branch periodic orbit at $Re=100$, $Wi=1.35$. Top: $c_{ii}$, symmetric logarithmic colour scale has been used except for a small region around 0. Bottom: $u_{2}$. Left: Mean flow, middle: real part fundamental oscillatory frequency, right: first harmonic. \label{fig:koopman_PO_emodes_ub}}
\end{figure}

\begin{figure}
\begin{center}
\includegraphics[width=0.75\textwidth]{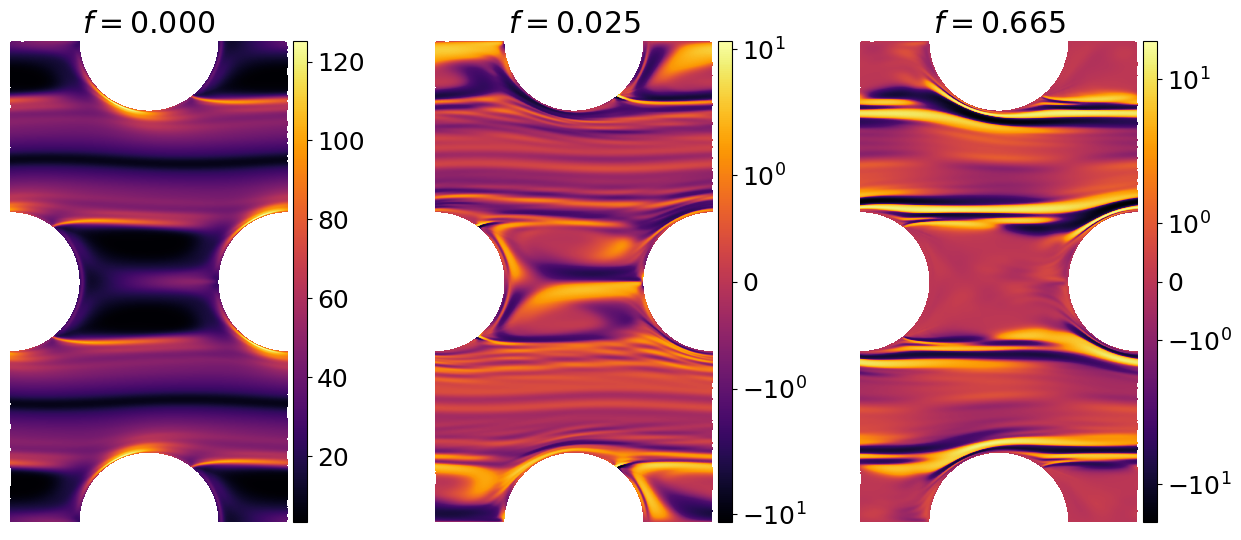}\\
\includegraphics[width=0.75\textwidth]{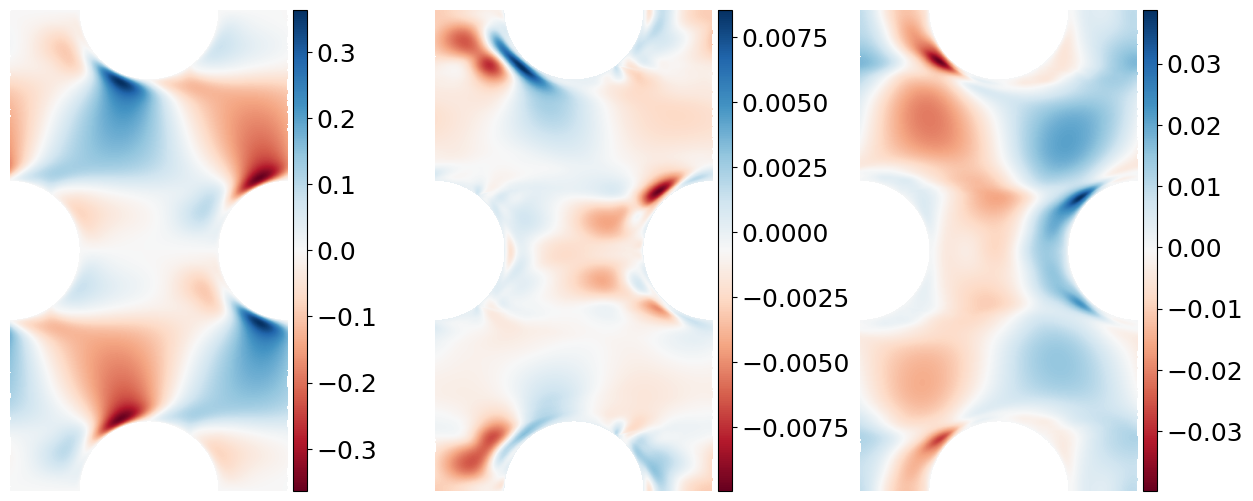}
\end{center}
\caption{Contours of the Koopman modes from EIT at $Re=100$, $Wi=10$. Top: $c_{ii}$, symmetric logarithmic colour scale has been used except for a small region around 0. Bottom: $u_{2}$. Left: Mean flow mode, middle: real part of a mode with frequency $f<f_0$, right: real part of a mode with frequency $f\geq f_0$, note the structural similarity between this mode with $f=0.665$ and the fundamental physical mode of the lower branch solution in figure~\ref{fig:koopman_PO_emodes_lb}.\label{fig:koopman_chaos_emodes}}
\end{figure}

We use ResDMD to study the upper and lower branches within the multistable region at $Re=100$, $Wi=1.35$, where two different stable periodic orbits coexist. We build our matrices $\bm{X}$ and $\bm{Y}$ using $M=80$ snapshots spaced by $\Delta t=0.05$ and choose a truncation of $r=21$ in the SVD. The leading eigenvalue in both periodic orbits is purely real $\lambda_0=1$, and corresponds to the mean flow. The subsequent eigenvalues come in complex conjugate pairs, and each pair represent a physical oscillatory mode. In the lower-branch solution the eigenvalue for the leading oscillatory modes is $\lambda^{\text{LB}}_1=0.9766\pm0.2152i$, and the remaining eigenvalues correspond to higher harmonics, as expected in a periodic orbit. \reva{We recall that the frequency of the orbit (which would correspond to the first frequency peak in the energy spectra -- any higher frequency peaks would correspond to higher harmonics) can be recovered from the eigenvalues of the Koopman operator as \begin{equation}\label{eq:Koopman_eig_to_f}f=\Im{(\log{(\lambda_i)})}/(2\pi\Delta t),\end{equation} }where $\Delta{t}$ is the interval between snapshots~\citep{bagheri2013koopman}. We ensure that the leading eigenvalues for the lower branch solution present a residual $\text{res}(\bm{g}_i^{\text{LB}},\lambda_i^{\text{LB}})\leq10^{-3},i\in\{1,\dots,4\}$ for the eigenvalue pairs. The computed eigenvalues and the pseudospectrum are shown in figure~\ref{fig:koopman_evals} (left). \reva{All eigenvalues lie on the unit circle with low error, and correspond to higher harmonics of a fundamental frequency, confirming we are on a periodic orbit.} Contours of the mean flow as well as the real part of fundamental oscillatory mode and the first harmonic are shown in figure~\ref{fig:koopman_PO_emodes_lb} (top) for $c_{ii}$ and figure~\ref{fig:koopman_PO_emodes_lb} (bottom) for the $u_{2}$ component of the velocity. \reva{These figures display the physical structure of the underlying dynamic mechanisms sustaining the periodic orbit.}

The eigenvalues and pseudospectrum for the upper branch stable periodic orbit at $Re=100$, $Wi=1.35$ are shown in figure~\ref{fig:koopman_evals} (middle). The leading oscillatory mode for the upper branch has a value $\lambda^{\text{UB}}_1=0.9800\pm0.1991i$, and the remaining computed eigenvalues correspond to higher harmonics.  We ensure that the leading eigenvalues for the upper branch solution present a residual $\text{res}(\bm{g}_i^{\text{UB}},\lambda_i^{\text{UB}})\leq10^{-3},i\in\{1,\dots,4\}$ for the eigenvalue pairs, as we did for the lower branch solution. Contours of the mean flow as well as the real part of fundamental oscillatory mode and the first harmonic are shown in figure~\ref{fig:koopman_PO_emodes_ub} (top) for $c_{ii}$  and figure~\ref{fig:koopman_PO_emodes_ub} (bottom) for the $u_{2}$ component of the velocity. 

The chaotic solution at $Re=100$, $Wi=10$ requires much more data to ensure low residuals of the eigenvalues and eigenmodes of the Koopman operator. In this case we use $M=1300$ snapshots spaced by $\Delta t=0.3$ and choose a truncation of $r=301$ in the SVD. We show the resulting eigenvalues of the Koopman operator and contours of the pseudospectrum in figure~\ref{fig:koopman_evals} (right). We observe that in contrast with the stable periodic orbits at $Re=100$, $Wi=1.35$, many eigenvalues present significant residuals and are distributed inside the unit circle. Figure~\ref{fig:koopman_chaos_emodes} (top) shows contours of $c_{ii}$ and figure~\ref{fig:koopman_chaos_emodes} (bottom) shows contours of the $u_{2}$ component of the velocity. We show three modes of the Koopman operator corresponding to the mean flow, a low frequency oscillation $f=0.025<f_0$ ($\lambda=0.9923+0.0472i$) and a frequency $f=0.665\approx f_0$ ($\lambda=0.3024+0.9185i$). \revb{All the shown eigenvalues in the chaotic state have a residual $\leq 0.26$}. 

We observe that the stable periodic orbit on the upper and lower branches (figures~\ref{fig:koopman_PO_emodes_lb} and~\ref{fig:koopman_PO_emodes_ub}) have similar fundamental oscillatory frequencies. The mean flow of the upper branch solution presents larger regions of the fluid which are stretched than the lower branch counterpart in the intercylinder region. This feature is also evident in the fundamental oscillatory frequency as well as the higher harmonics. Studying the fundamental oscillatory frequency, we can observe that the upper branch presents thicker layers of polymer stretch in between the channels, leaving a smaller region of fluid with small polymer stretching. We notice for both the upper and lower branch solutions that the fundamental oscillatory mode is symmetric on the trace of the polymers, while the first harmonic is antisymmetric (which is reversed for the vertical velocity). It is worth mentioning that the antisymmetric structure of the polymer stretching in the channel-like regions in between cylinders suggests that arrowheads cannot be sustained within this flow. A further difference between the upper and lower branch solutions can be observed in the vertical velocity at the upstream point of the cylinder, which is more significant on the lower branch solution than the upper one. In contrast, the upper branch solution presents larger values of $u_2$ at the sides of the upstream point of the cylinder. We also observe that $u_2$ in the intracylinder region of the lower branch presents a smaller lengthscale than the upper branch. The oscillations of the cylinder wake are not significant in any of these solutions. Whilst we did not see any traces of PDI in our flow snapshots, we note that traces of PDI \citep{beneitez_2023} were observed in some of the higher harmonics on the periodic orbit solutions. However, these modes represent higher frequencies than the fundamental modes, with amplitudes an order of magnitude smaller. Furthermore, the transition to EIT involves the emergence of higher energy, low-frequency dynamics. We therefore consider that PDI does not drive the observed dynamics of either the periodic orbits or the transition to chaos.

We can compare the Koopman modes for the stable periodic orbits to the ones for EIT. Firstly, we see that the fundamental oscillatory frequency mode in figure~\ref{fig:koopman_PO_emodes_lb} presents a similar frequency and structure to the mode shown in figure~\ref{fig:koopman_chaos_emodes} (right), which is further evidence that EIT stems from several bifurcations of the lower branch periodic orbit, as discussed above. Moreover, in EIT an additional low-frequency mode emerges with a significantly different structure to those observed on the periodic orbits. This mode is shown in figure~\ref{fig:koopman_chaos_emodes} (center) and presents significantly stretched polymers in the channel-like regions between the cylinders, while also clearly highlighting cylinder wake oscillations, which are almost not present if $f\geq f_0$. \reva{The use of Koopman analysis in the present problem allows us to identify the physical mechanisms behind the transition to EIT. We observe that although apparently similar, the upper and lower branch have significantly different physical structure, as shown by the fundamental frequency mode from our analysis. We can then identify that the same physical structure corresponding to the fundamental frequency mode of the lower branch survives all the way into EIT. These results provide support for the identified RTN route giving rise to EIT.}

\subsection{Transition via increasing $Re$}\label{sec:wi10}

\begin{figure}
\setlength{\unitlength}{1cm}
\begin{center}
\begin{picture}(14,5)(0,0)
 \put(0,0){\includegraphics[width=0.49\textwidth]{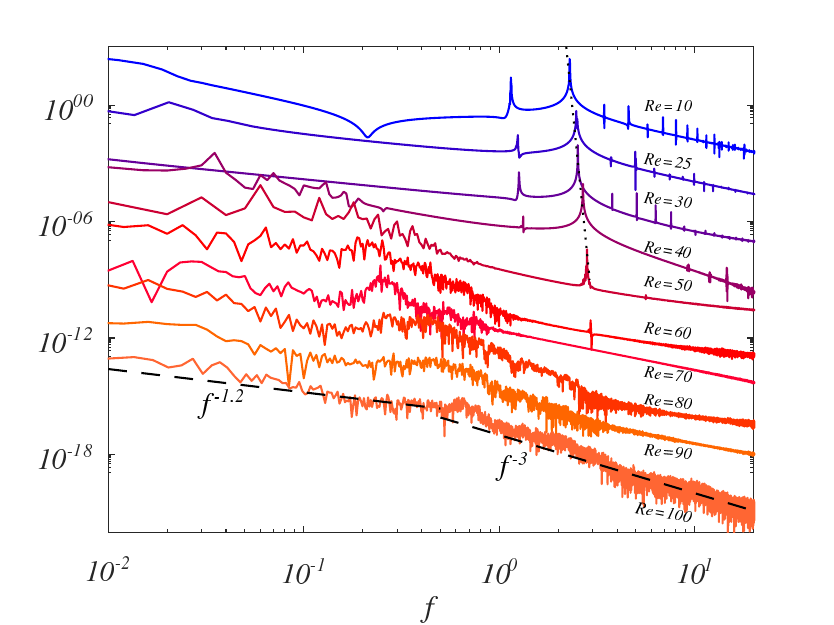}}
 \put(7,0){\includegraphics[width=0.49\textwidth]{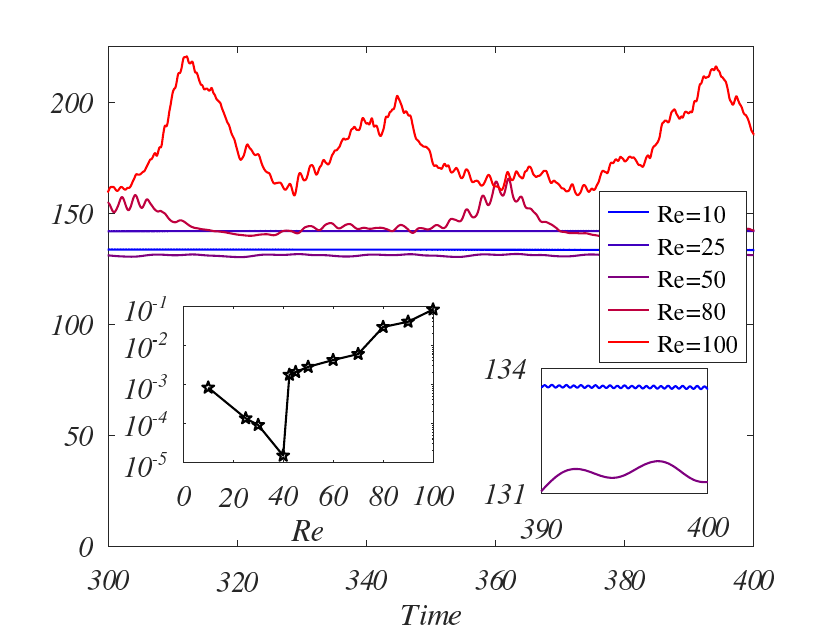}}
 \put(0,2.5){\rotatebox{90}{\scriptsize{$\widetilde{E_{K}}$}}}    
 \put(7,2.2){\rotatebox{90}{\scriptsize{$E_{E}/Wi$}}}    
 \put(8.45,2.65){\tiny{$\sigma_{t}\left(E_{E}\right)/\left\langle{E}_{E}\right\rangle_{t}$}}
 
 \end{picture}
\end{center}
\caption{Left panel: Frequency spectrum of the spatially averaged kinetic energy $\widetilde{E_{K}}$ for $Wi=10$ and a range of $Re$. Note the lines have been vertically shifted to allow for ease of visualisation. The black dotted line indicates the signature of arrowheads. Right panel: Time evolution of the volume averaged conformation tensor trace $E_{E}$ for several values of $Re$ at $Wi=10$. The left inset shows variation the normalised standard deviation $\sigma\left(E_{E}\right)/\left\langle{E}_{E}\right\rangle_{t}$ with $Re$, and the right inset shows a close up of $E_{E}$ for $Re=10$ and $Re=50$.\label{fig:wi10_f}}
\end{figure}

We next fix $Wi=10$ and vary $Re\in\left[10,100\right]$. The left panel of figure~\ref{fig:wi10_f} shows the energy spectra $\widetilde{E_{K}}$ for $Wi=10$ over this range of $Re$. The right panel of figure~\ref{fig:wi10_f} shows the time evolution of $E_{E}$ for $Wi=10$ and a range of $Re$. For $Re\le25$ we see clear arrowhead structures as visible in figure~\ref{fig:varRe_trc}, and the imprint of these in the kinetic energy spectra is clear in figure~\ref{fig:wi10_f} as the peak (connected by a dotted black line across different $Re$) with a frequency $f_{a}\approx2.1$ for $Re=10$, increasing to $f_{a}\approx{3}$ at $Re=50$. Note that the amplitude of this frequency peak decreases with increasing $Re$, as does the amplitude of a sub-harmonic at $f_{a}/2$. For $Re\ge50$, a large-amplitude, low-frequency signal dominates, and we do not observe any arrowhead structures in the flow snapshots. For $Re\ge60$, the imprint of the arrowhead in the energy spectra is not present.

The low-frequency signal visible in the left panel of figure~\ref{fig:wi10_f} for $Re\ge50$ corresponds to oscillations of the recirculating vortices in the wake, a snapshot of which can be seen in panel c) of figure~\ref{fig:varRe_trc}. At $Re=50$, we do not observe a power law spectra of $\widetilde{E_{K}}$ in figure~\ref{fig:wi10_f}; instead there are distinct peaks in the frequency spectrum, although the magnitude of these peaks scale with approximately $f^{-3}$ - the same slope as the power-law spectra for EIT at larger $Re$. As $Re$ is further increased we see a power-law spectra, with slope approximately $f^{-1.2}$ at low frequencies, and $f^{-3}$ at high frequencies. 

In the inset of figure~\ref{fig:wi10_f} we plot the standard deviation of the volume-averaged conformation tensor trace (normalised by the mean) $\sigma\left(E_{E}\right)/\left\langle{E}_{E}\right\rangle_{t}$ as a function of $Re$. For $Re\le40$, $\sigma\left(E_{E}\right)/\left\langle{E}_{E}\right\rangle_{t}$ decreases with increasing $Re$, by approximately an order of magnitude per doubling of $Re$. In this regime, the temporal fluctuations in the flow are driven by arrowhead structures, providing clear evidence that these structures weaken with increasing $Re$. It may be that the length of arrowheads increases with increasing $Re$, and one explanation is that the periodicity we impose by restricting our domain to the minimal repeating unit of the porous array. Alternatively, it could be that the interaction between wake oscillations/vortex shedding and the flow in the channels interferes with arrowheads, or prevents their development. However, with the focus of the present work on the transition to EIT, we do not investigate the weakening of the arrowhead structures further here. At $Re=42.5$ there is a transition to wake-oscillation dominated flow - the arrowheads may still be present (and their signal is visible in the spectra at $Re=50$ in the left panel of figure~\ref{fig:wi10_f}), but the flow dynamics is dominated by a lower frequency oscillation of the recirculation vortices in the cylinder wake. This transition is sub-critical, and can be traced back to $Re=40$. Regardless of the mechanism by which the arrowhead structures are suppressed at larger $Re$ in the current geometry, the data in the inset of the right panel of figure~\ref{fig:wi10_f} suggest that the onset of EIT here is related to the inertial mechanism of a wake instability leading to vortex shedding\revb{. Whilst we cannot rule out the existence of purely elastic instabilities, the separation of the unsteady dynamics at low $Re$ (arrowheads) and higher $Re$ (wake oscillations and vortex shedding), imply inertia plays a key role in the observed dynamics.}

\begin{figure}
\setlength{\unitlength}{1cm}
\begin{center}
\begin{picture}(14,5)(0,0)
 \put(0,0){\includegraphics[width=0.49\textwidth]{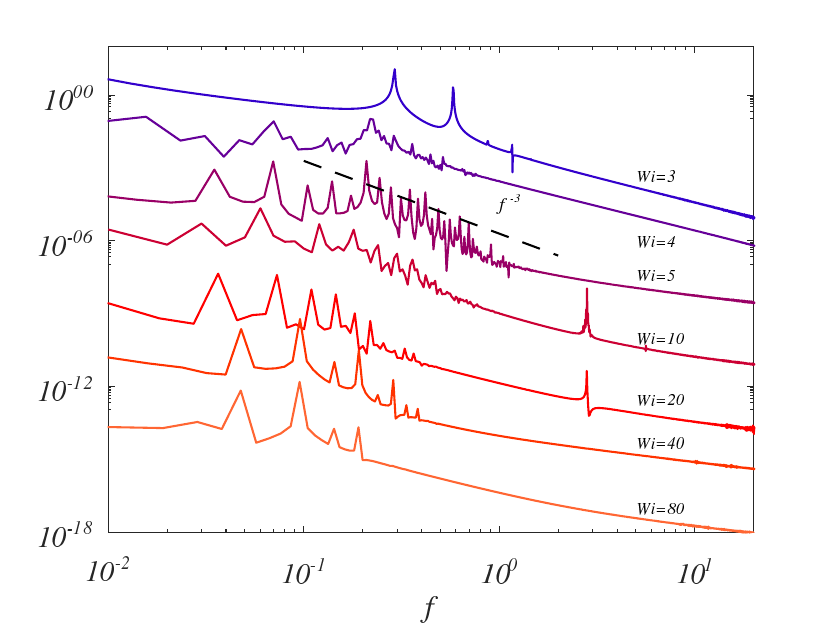}}
 \put(7,0){\includegraphics[width=0.49\textwidth]{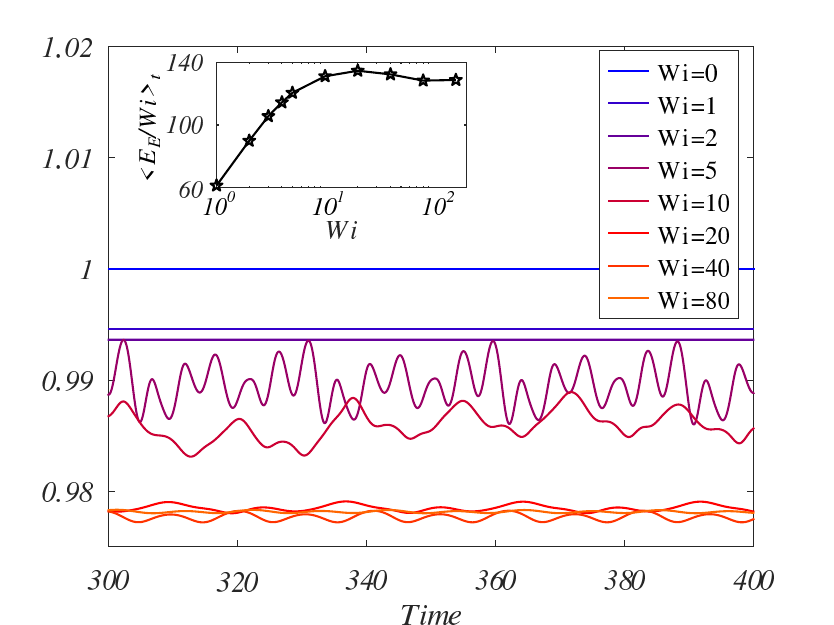}}
 \put(0,2.5){\rotatebox{90}{\scriptsize{$\widetilde{E_{K}}$}}}    
 \put(7,2.2){\rotatebox{90}{\scriptsize{$\dot{Q}$}}}    
 %\put(8.45,2.65){\tiny{$\sigma_{t}\left(E_{E}\right)/\left\langle{E}_{E}\right\rangle_{t}$}}
 
 \end{picture}
\end{center}
\caption{Left panel: Frequency spectrum of the spatially averaged kinetic energy $\widetilde{E_{K}}$ for $Re=50$ and a range of $Wi$. Note the lines have been vertically shifted to allow for ease of visualisation. Right panel: Time evolution of the volumetric flow rate $\dot{Q}$ for several values of $Wi$ at $Re=50$. The inset shows the variation with $Wi$ of $\left\langle{E}_{E}/Wi\right\rangle_{t}$ for $Re=50$.\label{fig:re50_f}}
\end{figure}

\begin{figure}
\begin{center}   
\includegraphics[width=0.99\textwidth]{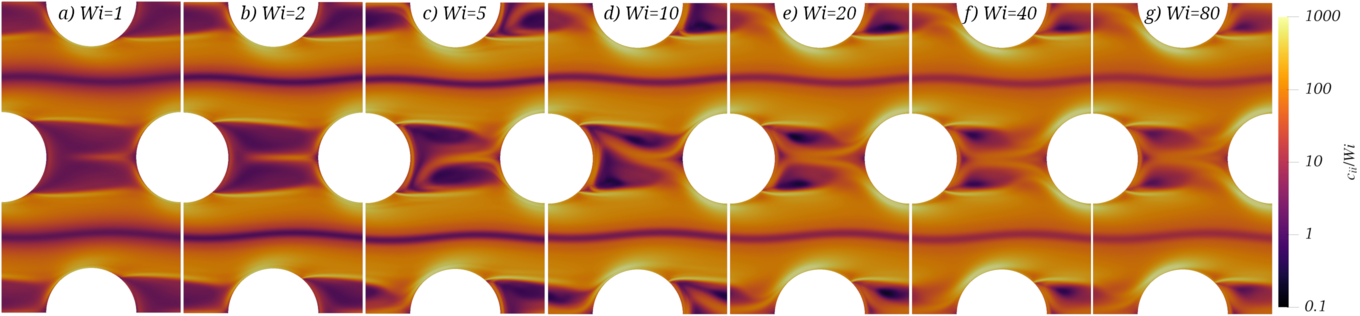}
\end{center}
\caption{Snapshots showing isocontours of the conformation tensor trace (normalised by $Wi$) for $Re=50$ and a range of $Wi$.\label{fig:re50_snapshots}}
\end{figure}

\subsection{$Re=50$ - between arrowheads and EIT}\label{sec:re50}

We next fix $Re=50$ and vary $Wi$. Figure~\ref{fig:re50_snapshots} shows instantaneous isocontours of $c_{ii}/Wi$ for $Re=50$ and a range of $Wi$. At all $Wi$ shown, the flow in the bulk - the channels between rows of cylinders - is primarily steady. For $Wi\le2$ the entire flow is steady. At larger $Wi$, oscillations of the recirculating vortex pair in the wake develop, with the strongest oscillations occurring between $Wi=5$ and $Wi=10$. The left panel of figure~\ref{fig:re50_f} shows the frequency spectrum of the kinetic energy for a range of $Wi\ge3$. We omit spectra at $Wi\le{2}$, as the flow is steady. As $Wi$ is increased, a range of frequencies develop, and for $Wi=5$ and $Wi=10$, the amplitudes of the peaks in $\widetilde{E_{K}}$ scale with approximately $f^{-3}$, although the spectra retain distinct peaks, rather than exhibiting a continuous power law. As we further increase $Wi\ge20$, we see a shift to fewer peaks in the spectra and these peaks are at lower frequencies. For $Wi=10$ and $Wi=20$, there is a small peak in $\widetilde{E_{K}}$ at $f_{a}\approx{3}$. This is the imprint of weak arrowhead structures, but is only present for $Wi\in\left[10,20\right]$, and not for smaller or larger values of $Wi$.

The right panel of figure~\ref{fig:re50_f} shows the time evolution of the volumetric flow rate $\dot{Q}$ for a range of $Wi$ at $Re=50$. There is a general trend of decreasing $\dot{Q}$ with increasing $Wi$, which saturates for $Wi\ge20$. For $Wi\le{2}$, the flow is steady, and at larger $Wi$ periodic behaviour is observed. Beyond $Wi=10$, the magnitude of the oscillations in $\dot{Q}$ decreases with increasing $Wi$, and the number of frequencies present reduces (to two). The inset of the right panel of figure~\ref{fig:re50_f} shows the variation of $\left\langle{E}_{E}/Wi\right\rangle_{t}$ with $Wi$, and we see an increase up to $Wi=10$, and a plateau - and slight decrease - at larger $Wi$. This is in contrast to the EIT case at $Re=100$ (inset of right panel of figure~\ref{fig:re100_t}) where $\left\langle{E}_{E}/Wi\right\rangle_{t}$ continues to grow with $Wi$ beyond the transition to chaotic dynamics. 

Whilst there is a weak signal of arrowheads present at $Re=50$ and $Wi=10,20$, the dynamics of these flows is clearly dominated by a different mechanism - the oscillations of the recirculating wake, which occur at much longer timescales here. Whilst the wake is steady in the Newtonian limit at $Re=50$, the wake oscillations at larger $Wi$ are not a purely elastic effect, as inertia is necessary for a separated flow (steady or unsteady). The absence of a chaotic flow state at $Re=50$ and all $Wi$ simulated further suggests that in the present configuration, the EIT state is disconnected from low-inertia states where purely elastic effects - and arrowheads - dominate the dynamics.

\section{Conclusions\label{sec:conc}}

We have numerically investigated the transition to elasto-inertial turbulence (EIT) in cylinder arrays, as a canonical representation of a porous media. We find that in the present configuration, the transition to EIT as elasticity increases occurs first via a subcritical saddle-node bifurcation to a low-elastic stress, high-kinetic energy state in which vortex shedding is partially suppressed, followed by a Ruelle-Takens-Newhouse type transition to chaos. Arrowhead structures do not play a role in the transition to EIT \emph{in the present configuration}. They exist at small $Re$, but are suppressed with increasing inertia. We observe a distinct two-slope power law energy spectra, with the slope changing at the vortex shedding frequency. Koopman analysis of our simulations reveals that this two-slope spectrum arises from two mechanisms at different time scales: vortex shedding triggers an EIT-like state in the channels between cylinders, whilst wake-oscillations influence the slower dynamics. For a fixed $Wi$ and increasing $Re$, arrowheads are suppressed, and the transition to chaotic dynamics occurs via a subcritical transition to wake oscillations, which interact with the flow in the channels above and below, leading to EIT. There is no connection (in the present configuration) between arrowhead-dominated dynamics and EIT, which is triggered via an inertial vortex-shedding instability interacting with flow in the channels between cylinders.

We have shown EIT may be triggered here by local pore-scale vortex shedding, but it is possible a connection to elastic turbulence (ET) and arrowhead structures may be made if larger-scale structures are permitted via a relaxation of the periodicity \revc{or by extending the domain beyond a single-unit cell}. \revc{Multi-pore structures have previously been observed in experiments, therefore an exploration of the effects of scale, although beyond the scope of this work, would assist in understanding this connection}. The two-dimensional chaotic dynamics observed in our simulations are distinct from inertial turbulence, occuring at significantly lower $Re$, and potentially providing a route to lower-energy mixing in these canonical porous geometries.

\section*{Acknowledgements}
JK is funded by the Royal Society via a University Research Fellowship (URF\textbackslash R1\textbackslash 221290). We would like to acknowledge the assistance given by Research IT and the use of the Computational Shared Facility at the University of Manchester. We thank EPSRC for computational time made available on the UK supercomputing facility ARCHER2 via the UK Turbulence Consortium EP/X035484/1). 

\section*{Declaration of interests}
The authors report no conflict of interest. 

\appendix

\section{Convergence of simulations}\label{sec:conv}

\begin{figure}
\setlength{\unitlength}{1cm}
\begin{center}
\begin{picture}(14,5)(0,0)
  \put(0,0){\includegraphics[width=0.49\textwidth]{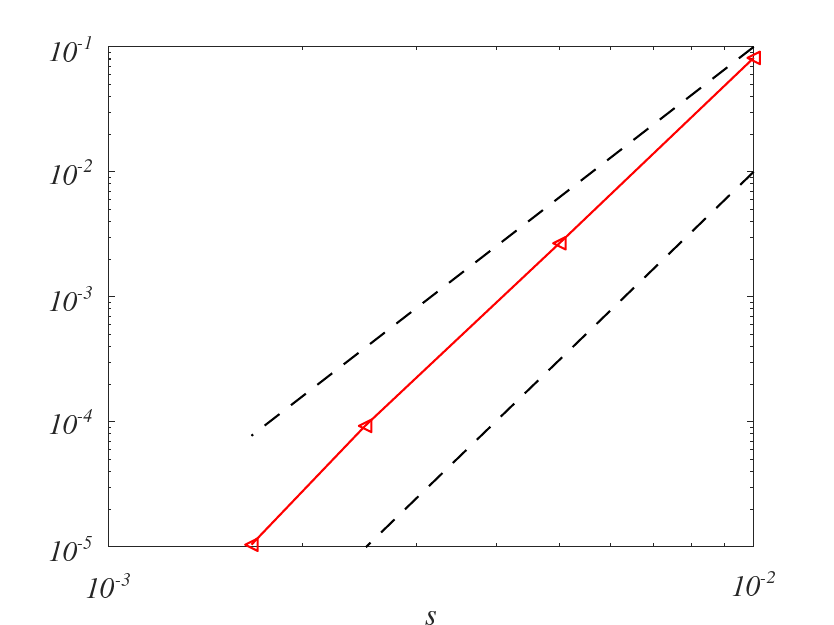}}
  \put(7,0){\includegraphics[width=0.49\textwidth]{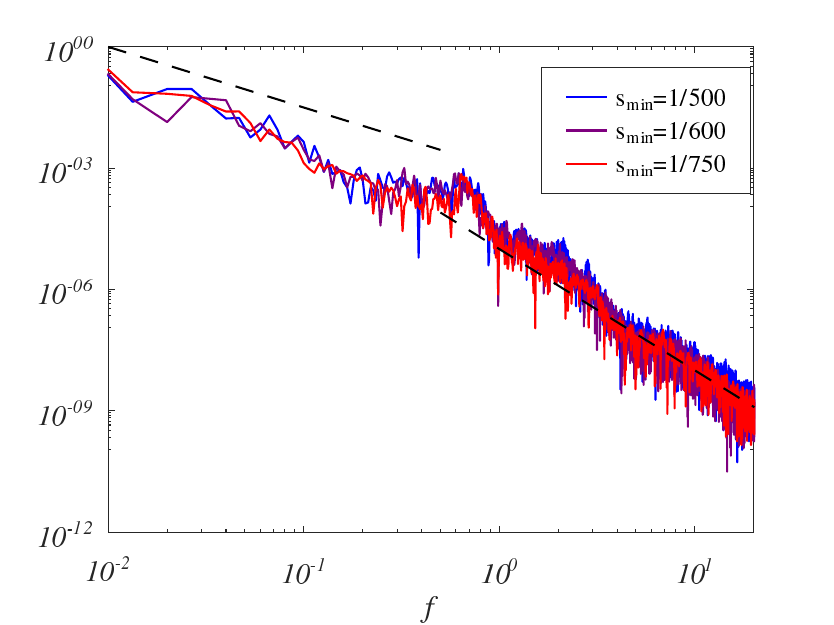}}
  \put(0,2.5){\rotatebox{90}{\scriptsize{$\lvert\dot{Q}-\dot{Q}^{\star}\rvert$}}}
  \put(2.6,2.5){\scriptsize{$\propto{s}^{4}$}}
  \put(4,1.5){\scriptsize{$\propto{s}^{5}$}}  
  \put(7.0,2.2){\rotatebox{90}{\scriptsize{$\widetilde{E_{K}}$}}}
\end{picture}
\end{center}
\caption{\reve{Left panel: Convergence of $\dot{Q}$ with resolution $s$ for $Re=50$, $Wi=1$. The dashed lines indicate slopes of $4$ and $5$. Right panel:} Frequency spectrum of the spatially averaged kinetic energy $\widetilde{E_{K}}$ at $Re=100$ and $Wi=10$ for three resolutions.\label{fig:res}}
\end{figure}

Here we provide verification that our simulations are converged with respect to both resolution and compressibility. \reve{The spatial discretision scheme used is sixth-order in the bulk, dropping to fourth-order at walls. Whilst the convergence properties of the method have been extensively analysed in~\cite{king_2020,king_2022,king_2024a}, we here provide confirmation that our results are converging in the setting of cylinder arrays. We first use a uniform resolution $s$ throughout the domain, and taking the case $Re=50$, $Wi=1$, for which the flow is steady, we plot in the left panel of figure~\ref{fig:res} the absolute error in volumetric flow rate $\dot{Q}$ with $s$. In lieu of an exact solution, we take the volumetric flow rate for a high resolution simulation $s=1/800$ as a reference, denoted $\dot{Q}^{\star}$. $\lvert\dot{Q}-\dot{Q}^{\star}\rvert$ scales with approximately $s^{4.3}$, lying between the expected bulk and boundary convergence rates.} The most challenging regions to resolve - with the finest flow structures - in the present geometry are the trailing stagnation/separation points on the cylinder surface. For all our simulations we set the finest resolution $s_{min}$ on the wall, increasing to $s_{max}=5s_{min}$ in the far field. Figure~\ref{fig:res} shows the frequency spectrum of the spatially averaged kinetic energy $\widetilde{E_{K}}$ at $Re=100$ and $Wi=10$ for three resolutions, $s_{min}\in\left[1/500,1/600,1/750\right]$. For all three resolutions, there is a \reva{close} agreement in the magnitude and slopes of the spectrum. We conducted additional simulations at $Re=100$ and $Wi\in\left[1,5\right]$ with $s_{min}=1/500$, and observed the transition pathway to EIT is the same as in our simulations with $s_{min}=1/600$. Additionally, for simulations at $Re=10$, $Wi=160$ for $s_{min}\in\left[1/400,1/500,1/600,1/750\right]$, we find that for $s_{min}\le1/500$ the flow dynamics (unsteady arrowheads) are independent of the resolution, and the mean kinetic energy $\left\langle{E}_{K}\right\rangle_{t}$ is converged to within $0.4\%$.

\begin{figure}
\setlength{\unitlength}{1cm}
\begin{center}
\begin{picture}(14,5)(0,0)
 \put(0,0){\includegraphics[width=0.49\textwidth]{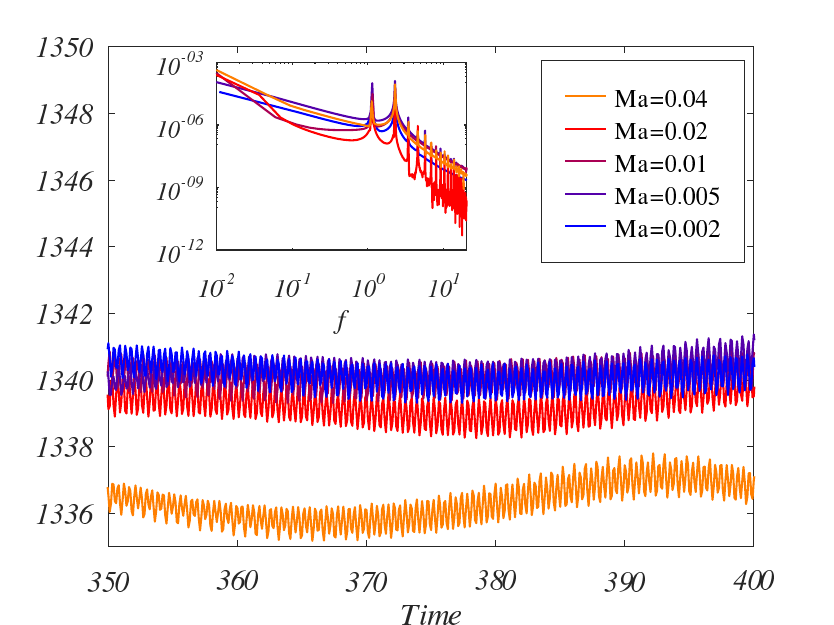}}
 \put(7,0){\includegraphics[width=0.49\textwidth]{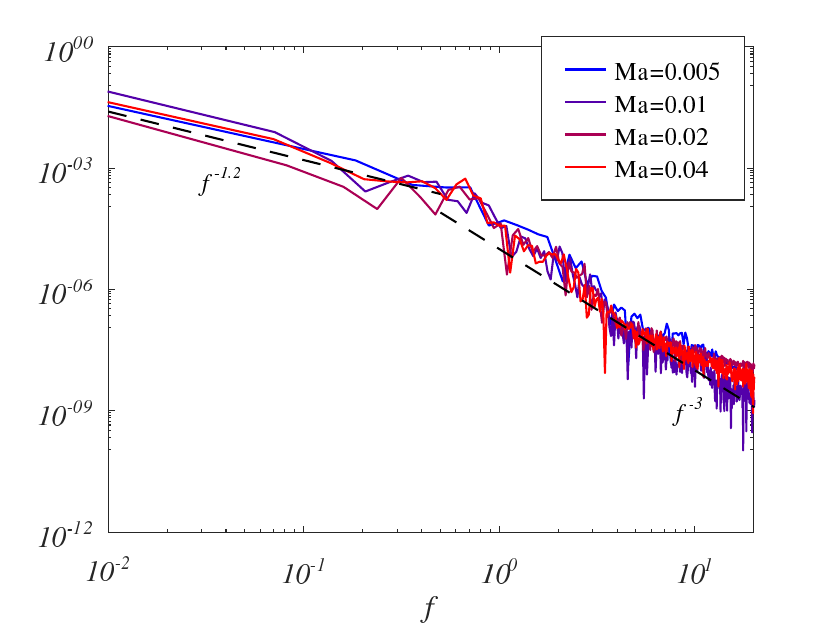}}
 \put(0.0,2.5){\rotatebox{90}{\scriptsize{$E_{E}$}}}    
 \put(7,2.5){\rotatebox{90}{\scriptsize{$\widetilde{E_{K}}$}}}
 \put(1.0,3.6){\scriptsize{$\widetilde{E_{K}}$}}
% \put(3.0,0){\scriptsize{\emph{Time}}} 
% \put(10.0,0){\scriptsize{\emph{Time}}} 
% \put(4.55,0.7){\tiny{\emph{Wi}}}
% \put(11.55,0.7){\tiny{\emph{Wi}}}     
\end{picture}
\end{center}
\caption{The influence of $Ma$ at $Wi=10$, for $Re=10$ (left panel) and $Re=100$ (right panel). The left panel shows the time-evolution of the volume averaged conformation tensor trace $E_{E}=\left\langle{c}_{ii}\right\rangle_{\mathcal{V}}$, with the inset showing the frequency spectrum of the volume averaged kinetic energy, for $Re=10$. The right panel shows the frequency spectrum of the volume averaged kinetic energy for $Re=100$.\label{fig:mach}}
\end{figure}

We next return to our standard resolution of $s_{min}=1/600$, and vary $Ma$, to ensure our results are not affected by compressibility. The left panel of figure~\ref{fig:mach} shows the time evolution of the volume averaged conformation tensor trace $E_{E}$ at $Re=10$, $Wi=10$ for $Ma\in\left[0.002, 0.005, 0.01, 0.02, 0.04\right]$, and the inset shows the frequency spectrum of the kinetic energy $\widetilde{E_{K}}$ for the same simulations. For all \reva{five} values of $Ma$, we see peaks in the spectra at the same frequencies, associated with the passage of arrowhead structures through the domain, and a similar trend at low frequencies due to interactions between arrowheads. For all three values of $Ma$, $\left\langle{E}_{E}\right\rangle_{t}$ is converged to within $0.35\%$. $Re=10$ is the smallest value of $Re$ studied in this work, and consequently that with the largest pressure gradient $F_{0}$; it is therefore the case where compressibility effects are largest. The right panel of figure~\ref{fig:mach} shows the frequency spectrum of the kinetic energy $\widetilde{E_{K}}$ for the same values of $Ma$ at $Re=100$, $Wi=10$, and we see \reva{close agreement between all $Ma$ tested in terms of the slopes of the spectra above and below the vortex shedding frequency. This suggests} that for $Re=100$, a Mach number of $0.04$ is sufficient the ensure compressibility effects may be neglected. However, to ensure consistency across our simulations at all $Re$, we choose to set $Ma=0.01$. \revb{At $Ma=0.01$, the $L_{2}$ norm of $\partial{u}_{i}/\partial{x}_{i}$ does not exceed $\mathcal{O}\left(10^{-4}\right)$.}

\bibliographystyle{jfm}
\bibliography{jrckbib}

\end{document}